\documentclass[12pt]{iopart}
\usepackage{amssymb}
\usepackage{iopams}
\usepackage{amsthm}
\usepackage{amscd}
\usepackage{amsfonts}
\usepackage{color}
\usepackage{amsbsy}
\catcode`\@=11
\renewcommand\footnoterule{
 \kern-3\p@
 \hrule\@width.4\columnwidth
 \kern2.6\p@}
\renewcommand\@makefntext[1]{
 \parindent 1em\noindent
 \hb@xt@1.8em{\hss$^{\@thefnmark}$)}\hspace{2pt}
 \footnotesize\rmfamily#1}
\def\@makefnmark{\hspace{.5pt}\hbox{$^{\@thefnmark}$
\hspace{-1pt})}} 
\catcode`\@=11
\renewcommand\footnoterule{
 \kern-3\p@
 \hrule\@width.4\columnwidth
 \kern2.6\p@}
\renewcommand\@makefntext[1]{
 \parindent 1em\noindent
 \hb@xt@1.8em{\hss$^{\@thefnmark}$)}\hspace{2pt}
 \footnotesize\rmfamily#1}
\def\@makefnmark{\hspace{.5pt}\hbox{$^{\@thefnmark}$
\hspace{-1pt}}} 

\newtheorem{theorem}{Theorem}[section]
\newtheorem{proposition}[theorem]{Proposition}
\newtheorem{cor}[theorem]{Corollary}
\newtheorem{lemma}[theorem]{Lemma}
\newtheorem{definition}[theorem]{Definition}
\newtheorem{corollary}[theorem]{Corollary}
\theoremstyle{definition}

\newtheorem{remark}[theorem]{Remark}

\begin{document}
\title[$S$-matrix of Schr\"{o}dinger Operator with Non-Symmetric Zero-Range
Potentials]{On $S$-matrix of Schr\"{o}dinger Operators with Non-Symmetric Zero-Range
Potentials}

\author{P. A.~Cojuhari${}^a$,\ A.~Grod${}^b$ and S. Kuzhel${}^c$}

\address{${}^a$\ ${}^c$
AGH University of Science and Technology, Department of Applied Mathematics,
30-059 Krakow, Poland \\
${}^b$\ Institute of Mathematics of the National
Academy of Sciences of Ukraine, Kiev, Ukraine}
\eads{\mailto{cojuhari@agh.edu.pl},\ \mailto{andriy.grod@yandex.ua},\
\mailto{kuzhel@mat.agh.edu.pl}}
\begin{abstract}
Non-self-adjoint Schr\"{o}dinger operators $A_{\mathfrak{T}}$ which correspond to non-symmetric zero-range
potentials are investigated. We show that various properties of $A_{\mathfrak{T}}$ (eigenvalues, exceptional points, spectral singularities and the property of similarity to a self-adjoint operator) are completely determined by poles of the corresponding $S$-matrix.
\end{abstract}


\maketitle
\section{Introduction}
 This work was inspirited by an intensive development of Pseudo-Hermitian Quantum Mechanics (PHQM) during last
 decades \cite{BE}. The key point of such theory is the employing of non-self-adjoint operators for the description of experimentally observable data. Briefly speaking, in order to
interpret a given non-self-adjoint operator $A$ in a Hilbert space $\mathfrak{H}$ as a
physically meaning Hamiltonian we have to check the reality of its spectrum and to prove
the existence of a new inner product that ensures the (hidden) self-adjointness of $A$.

As usual, in PHQM studies, a non-self-adjoint operator $A$ admits a
representation $A=A_0+V$, where $A_0$ is a fixed (unperturbed) self-adjoint operator in $\mathfrak{H}$
and a non-symmetric potential $V$ is characterized by a set $\Upsilon=\{\varepsilon\}$ of complex parameters
$\varepsilon$.
One of important problems is the
description of quantitative and qualitative changes of spectrum
$\sigma(A)$ when $\varepsilon$ runs $\Upsilon$.
A typical evolution of properties is the following:
\begin{equation}\label{eue1}\fl
 \begin{array}{c}
 I \\
 \mbox{non-real} \\ \mbox{eigenvalues} \end{array}
   \  \leftrightarrow \ \begin{array}{c}
   II \\
   \mbox{spectral singularities} \\ \mbox{exceptional points} \end{array} \  \leftrightarrow  \
   \begin{array}{c}
   III \\
   \mbox{similarity to} \\ \mbox{a self-adjoint operator} \end{array}
\end{equation}

The properties of operators from domains I and III are quite obvious. In particular, the existence of non-real eigenvalues
means that $A$ cannot be realized as a self-adjoint operator for any choice of inner product.
On the other hand, the similarity property shows that $A$ turns out to be self-adjoint with respect to
a new inner product of $\mathfrak{H}$ which is equivalent to the initial one.
The domain II can be interpreted as a boundary between I and III and the corresponding operators will
keep only part of properties of I and III. For instance, if $A$ corresponds to II, then its spectrum is real
(similarly to III) but $A$ cannot be made self-adjoint by an appropriative choice of equivalent inner product of $\mathfrak{H}$ (in spirit of
I).
This phenomenon deals with the appearing of `wrong' spectral points of $A$ which are impossible for the spectra
of self-adjoint operators. Traditionally, these spectral points are called \emph{exceptional points} if they are located at the
discrete spectrum of $A$ and \emph{spectral singularities} in the case of the continuous spectrum.
Exceptional points correspond to situations where two or more eigenvalues together with their eigenvectors coalesce.
Similar interpretation can also be carried out for spectral singularities with the use of generalized eigenvectors corresponding to
the continuous spectrum. The presence of exceptional points/spectral singularities indicate that we lose
the completeness of eigenvectors corresponding to eigenvalues and the continuous spectrum.
Nowadays, various aspects of exceptional points/spectral singularities including the physical meaning and possible practical
applications has been analyzed with a wealth of technical tools (see, e.g.,
\cite{GU} for exceptional points and \cite{MOST} for spectral singularities).

In the present paper, we show that the evolution of spectral properties (\ref{eue1})
can be successfully and easily described in terms of poles of $S$-matrices of operators $A$.
We illustrate this point by considering the set of operators $A$ generated by
the Schr\"{o}dinger type differential expression $A=-\frac{d^2}{dx^2}+V$ with zero-range potential
$$
V=a<\delta,\cdot>\delta(x)+b<\delta',\cdot>\delta(x)+
 c<\delta,\cdot>\delta'(x)+d<\delta',\cdot>\delta'(x),
$$
where $\delta$ and $\delta'$ are, respectively, the Dirac $\delta$-function
and its derivative and $a,b,c,d$ are complex numbers.

The Schr\"{o}dinger operator with zero-range potential fits well the Lax-Phillips scattering scheme
\cite{LF} since the potential is concentrated at one point (so-called $0$-perturbations \cite{operturb}).
In that case the $S$-matrix (the Lax-Phillips scattering matrix) of $A$ coincides with the meromorphic matrix-valued function
\begin{equation}\label{red1b}\fl
{\textsf S}(k)=[\sigma_0-2(1-ik)\mathfrak{T}][\sigma_0-2(1+ik)\mathfrak{T}]^{-1}, \quad k\in\mathbb{C}, \quad \sigma_0=\left(\begin{array}{cc}
 1 & 0 \\
 0 & 1
 \end{array}\right)
\end{equation}
where $(2\times{2})$-matrix $\mathfrak{T}$ is expressed in terms of parameters $a,b,c,d$ and it
determines the domain of definition of $A$ (see (\ref{sese2c})).
If the matrix $\mathfrak{T}$ is \emph{Hermitian}, then the corresponding operator $A=A_{\mathfrak{T}}$ is self-adjoint and
the $S$-matrix (\ref{red1b}) is the direct consequence of mathematically rigorous arguments of scattering theory:
establishing the existence of wave operators with subsequent representation of the scattering operator in the spectral
representation of unperturbed dynamics \cite{KUOLD}.

In the case of a non-self-adjoint operator $A_{\mathfrak{T}}$
we define the $S$-matrix by analogy, considering \emph{an arbitrary matrix} $\mathfrak{T}$ in (\ref{red1b}) and do not take
care about auxiliary mathematical things (see \cite{NEW}, \cite{CK} for details).
We found this definition useful because: a) the formula (\ref{red1b}) for $S$-matrix enables
to determine explicitly the matrix  $\mathfrak{T}$ characterizing the operator $A_{\mathfrak{T}}$;
b) the formula (\ref{red1b}) can be easily rewritten in terms of right/left reflection and transmission coefficients
of the corresponding traveling wave functions (cf. subsection 2.4).

The Lax-Phillips scattering scheme allows to define $S$-matrices
for Schr\"{o}dinger operators with local (i.e., the support of potential is a bounded interval)
non-symmetric potentials. The obtained formulas are similar to (\ref{red1b}) and they also can be rewritten via
reflection/transmission coefficients \cite{CK}. We believe that
such an interpretation of $S$-matrix which comes from the Lax-Phillips scattering theory
makes it possible to establish more informative connection between poles of $S$-matrix and spectral properties of
Schr\"{o}dinger operators with local non-symmetric potentials.

In this paper, using the decomposition of the $S$-matrix (\ref{red1b}) with respect to the Pauli matrices
(subsection 2.3), we show that the location of poles of the $S$-matrix ${\textsf S}(\cdot)$ completely determines
the spectral properties of non-self-adjoint operators $A_{\mathfrak{T}}$ outlined in (\ref{eue1}).

Our proof of similarity of $A_{\mathfrak{T}}$ to a self-adjoint operator in section 3 does not contain an algorithm of the construction
of an appropriative metric operator $e^Q$ which guarantees the self-adjointness of $A_{\mathfrak{T}}$. However, in the particular case
where the $S$-matrix of a non-self-adjoint operator $A_{\mathfrak{T}}$ has simple imaginary poles, we `guess' an explicit expression of the metric operator
(section 4).  Sections 5 and 6 are devoted to spectral singularities and exceptional points, respectively.

Throughout the paper $\mathcal{D}(A)$ denotes the domain and $\ker{A}$ denotes the null-space of a
linear operator $A$.  The resolvent set and the spectrum of $A$
are denoted by $\rho(A)$ and $\sigma(A)$, respectively.

\setcounter{equation}{0}
\section{Schr\"{o}dinger operator with non-symmetric zero-range potentials}
\label{sec2}
\subsection{Preliminaries}
A one-dimensional Schr\"{o}dinger operator with general zero-range potential at the point $x=0$ can be defined by
the formal expression
 \begin{equation}\label{lesia11}\fl
 -\frac{d^2}{dx^2}+a<\delta,\cdot>\delta(x)+b<\delta',\cdot>\delta(x)+
 c<\delta,\cdot>\delta'(x)+d<\delta',\cdot>\delta'(x),
 \end{equation}
where $\delta$ and $\delta'$ are, respectively, the Dirac $\delta$-function
and its derivative (with support at $0$) and $a,b,c,d$ are complex
numbers. Using the regularization method suggested in  \cite{AL1},  a direct relationship
between parameters $a, b, c, d$ of the singular potential
\begin{equation}\label{gu14}\fl
V=a<\delta,\cdot>\delta(x)+b<\delta',\cdot>\delta(x)+
 c<\delta,\cdot>\delta'(x)+d<\delta',\cdot>\delta'(x)
\end{equation}
and operator-realizations of (\ref{lesia11}) in the Hilbert space $L_2(\mathbb{R})$ can be established \cite{AlKuz}.
Precisely, the formal expression (\ref{lesia11}) gives rise to operators
\begin{equation}\label{lesia700}
A_{\mathbf{T}}=-\frac{d^2}{dx^2}, \qquad \mathbf{T}=\left(\begin{array}{cc}
 a & b \\
 c & d
 \end{array}\right).
\end{equation}
defined on smooth functions $f\in{W}_2^2(\mathbb{R}\backslash\{0\})$
which satisfy the boundary condition
 \begin{equation}\label{lesia800}
 \mathbf{T}\left(\begin{array}{c}
 \displaystyle{\frac{f(0+)+f(0-)}{2}} \vspace{3mm} \\
\displaystyle{\frac{-f'(0+)-f'(0-)}{2}}
\end{array}\right)=\left(\begin{array}{c}
f'(0+)-f'(0-) \vspace{3mm} \\
f(0+)-f(0-)
\end{array}\right).
 \end{equation}

\begin{remark}
The matrix $\mathbf{T}$ in (\ref{lesia800}) relates
the mean values of functions $f, f'$  at point $0$ with
their jumps.  Another description of point interaction at point
$0$ can be given by the matching conditions
\begin{equation}\label{gu2}
\mathbf{B}\left(\begin{array}{c}
f(0-) \\
f'(0-)
\end{array}
\right)=\left(\begin{array}{c}
f(0+) \\
f'(0+)
\end{array}
\right)
\end{equation}
which connect the left-side and the right-side boundary values of the functions $f, f'$ at point $0$ \cite{AlKur}.
The sets of operators defined via the boundary conditions (\ref{lesia800}) and (\ref{gu2}) do not coincide.
For instance, the operator $A_{\mathbf{T}}$ with ${\mathbf{T}}=\left(\begin{array}{cc}
0 & -2 \\
2 & 0 \end{array}\right)$ cannot be realized with the use of (\ref{gu2}).
\end{remark}

\subsection{Definition and elementary properties of $S$-matrix}
The $S$-matrix ${\textsf S}(\cdot)$ of
$A_{\mathbf{T}}$  can be directly expressed in terms of $\mathbf{T}$ since the potential $V$ is supported at point $0$.
However, the obtained formula looks quite cumbersome.
Having in mind to simplify the expression for  ${\textsf S}(\cdot)$, we rewrite the boundary condition (\ref{lesia800})
in the form
\begin{equation}\label{sese2c}
\mathfrak{T}\left(\begin{array}{c}
f(0+)+f'(0+) \\
f(0-)-f'(0-)
\end{array}\right)=\frac{1}{2}\left(\begin{array}{c}
f(0+) \\
f(0-)
\end{array}\right), \quad \mathfrak{T}=\left(\begin{array}{cc}
 \mathfrak{t}_{11} & \mathfrak{t}_{12} \\
 \mathfrak{t}_{21} & \mathfrak{t}_{22}
 \end{array}\right).
\end{equation}

It should be noted that the set of operators $A_{\mathfrak{T}}=-\frac{d^2}{dx^2}$  determined by the boundary condition
(\ref{sese2c}) does not coincide with the set of operators $A_{\mathbf{T}}$ defined by (\ref{lesia700}), (\ref{lesia800}).
Namely, the domain of definition of $A_{\mathbf{T}}$ admits
the presentation (\ref{sese2c}) if and only $-1$ does not belong to the point spectrum of  $A_{\mathbf{T}}$
or, that is equivalent (see \cite{GROD}), if
\begin{equation}\label{gu4}
\Xi=4-\mathbf{{det} \ T} + 2(a-d)\not=0,  \qquad \mathbf{{det} \ T}=ad-bc.
\end{equation}
In that case, it is easy to verify by comparing  (\ref{lesia800}) and (\ref{sese2c}) that
\begin{equation}\label{gu3}
\mathfrak{T}=\frac{1}{4\Xi}\left(\begin{array}{cc}
\Xi+2(b+c-a-d) & 4+\mathbf{{det} \ T}-2(b-c) \\
4+\mathbf{{det} \ T}+2(b-c)  & \Xi-2(b+c+a+d)
\end{array}\right).
\end{equation}
On the other hand, not every operator $A_{\mathfrak{T}}$ can be rewritten as $A_{\mathbf{T}}$
(for example $A_{\mathfrak{T}}$ with ${\mathfrak{T}}=0$ does not belong to the set of operators $A_{\mathbf{T}}$).

The operators $A_{\mathfrak{T}}$ fit well the Lax-Phillips scattering scheme and
the corresponding $S$-matrix of $A_{\mathfrak{T}}$  coincides with the matrix-valued function
\begin{equation}\label{red1}
{\textsf S}(k)=[\sigma_0-2(1-ik)\mathfrak{T}][\sigma_0-2(1+ik)\mathfrak{T}]^{-1}, \quad \sigma_0=\left(\begin{array}{cc}
 1 & 0 \\
 0 & 1
 \end{array}\right)
\end{equation}
determined for all $k\in\mathbb{C}_+=\{k\in\mathbb{C} \ :  \ \textsf{Im} \ k>0\}$  where (\ref{red1}) is well posed \cite{NEW}, \cite{CK}.

The expression (\ref{red1}) enables to determine the $S$-matrix ${\textsf S}(\cdot)$ of $A_{\mathfrak{T}}$ for any $(2\times{2})$-matrix $\mathfrak{T}$.
In the particular case where $\mathfrak{T}$ admits the representation (\ref{gu3})
(i.e., the matrix $\mathfrak{T}$ can be expressed via ${\mathbf{T}}$ and hence,
$A_{\mathfrak{T}}\equiv{A_{\mathbf{T}}}$) \emph{we will say that ${\textsf S}(\cdot)$ is the $S$-matrix of $A_{\mathbf{T}}$}.

\begin{remark}\label{rrr4}
\begin{enumerate}
  \item In the Lax-Phillips scattering scheme the free evolution is determined
  by the Friedrichs extension of the symmetric operator
  \begin{equation}\label{new9}\fl
  A_{s}=-\frac{d^2}{dx^2}, \qquad \mathcal{D}(A_{s})=\left\{f\in{W}_2^2(\mathbb{R}\backslash\{0\}) \ : \ \begin{array}{l}
  f(0+)=f(0-)=0 \\
  f'(0+)=f'(0-)=0
  \end{array}\right\}
  \end{equation}
  associated with given differential expression (\ref{lesia11}).
  Namely, the Friedrichs extension coincides with the operator
  \begin{equation}\label{new5}\fl
  A_F=-\frac{d^2}{dx^2}, \qquad \mathcal{D}(A_F)=\{f\in{W}_2^2(\mathbb{R}\backslash\{0\}) \ : \ f(0+)=f(0-)=0\}.
  \end{equation}
  The self-adjoint operator $A_F$ is determined by $\mathfrak{T}=0$ in (\ref{sese2c}).
  Thus, the matrix $\mathfrak{T}$ characterizes `a deviation'
  of $A_{\mathfrak{T}}$ from the unperturbed Hamiltonian $A_F$. In some sense, this
  explains why the matrix $\mathfrak{T}$ (rather than $\mathbf{T}$) appears in (\ref{red1}).
\item
The self-adjoint operator
\begin{equation}\label{new12}\fl
A_K=-\frac{d^2}{dx^2}, \qquad \mathcal{D}(A_K)=\{f\in{W}_2^2(\mathbb{R}\backslash\{0\}) \ : \ f'(0+)=f'(0-)=0\}
\end{equation}
 is determined by $\mathfrak{T}=\frac{1}{2}\sigma_0$ in (\ref{sese2c}) and it is the Krein extension
 of the symmetric operator $A_s$.  Similarly to the Friedrichs extension $A_F$, the Krein extension $A_K$ determines
 a free evolution in the Lax-Phillips scattering scheme \cite{MATNOTES}.
 The corresponding $S$-matrices are: ${\textsf S}(k)=\sigma_0$ for $A_F$ and
${\textsf S}(k)=-\sigma_0$ for $A_K$.
 \item The expression (\ref{red1}) determines the $S$-matrix for $A_{\mathbf{T}}$
  only in the case where $-1\in\rho(A_{\mathbf{T}})$.
It turns out that the formula (\ref{red1}) and the results below
can be easily modified for any operator $A_{\mathbf{T}}$ with nonempty resolvent set.
\end{enumerate}
\end{remark}

\smallskip

It follows from (\ref{red1}) that the $S$-matrix ${\textsf S}(\cdot)$ is a meromorphic matrix-function on $\mathbb{C}_+$.
It can be established that poles of ${\textsf S}(\cdot)$ correspond to eigenvalues of
$A_{\mathfrak{T}}$. Precisely, \emph{$k\in\mathbb{C}_+$ is a pole of ${\textsf S}(\cdot)$ if and only if
$k^2$ is an eigenvalue of $A_{\mathfrak{T}}$} \cite{GROD}.
The formula (\ref{red1}) allows to extend the definition of $S$-matrix of $A_{\mathfrak{T}}$ to all complex numbers $k\in\mathbb{C}$
satisfying the condition $\det[\sigma_0-2(1+ik)\mathfrak{T}]\not=0$.  Obviously, the extended $S$-matrix
remains to be a meromorphic matrix-function.

We will say that ${\textsf S}(\cdot)$ has \emph{a pole at infinity} if at least one of entries of ${\textsf S}(k)$ tend to infinity when $k\to\infty$. We will say that ${\textsf S}(k)$ is defined on the \emph{physical sheet} if
$k\in\mathbb{C_+}$  and ${\textsf S}(k)$ is defined on the \emph{nonphysical sheet} if $k\in\mathbb{C_-}=\{k\in\mathbb{C} \ :  \ \textsf{Im} \ k<0\}$.

According to the above discussion \emph{the discrete spectrum of $A_{\mathfrak{T}}$ is
completely determined by the corresponding $S$-matrix on the physical sheet $\mathbb{C}_+$}.

\subsection{The presentations of $S$-matrix with the use of Pauli matrices}
The $S$-matrix for a non-self-adjoint operator $A_{\mathfrak{T}}$ may have new unusual properties. For this reason,
an additional representations of ${\textsf S}(\cdot)$
can be useful. First of all, we are interesting in the decomposition of  ${\textsf S}(\cdot)$ with respect
to the Pauli matrices
$$
\sigma_0=\left(\begin{array}{cc}
1 & 0 \\
0 & 1 \end{array}\right), \quad
\sigma_1=\left(\begin{array}{cc}
0 & 1 \\
1 & 0 \end{array}\right), \quad \sigma_2=\left(\begin{array}{cc}
0 & -i \\
i & 0 \end{array}\right), \quad \sigma_3=\left(\begin{array}{cc}
1 & 0 \\
0 & -1 \end{array}\right).
$$

Let $X$ be an arbitrary $(2\times{2})$-matrix. Then $X$ admits the presentation
 $X=\sum_{j=0}^3{x_j}\sigma_j$, where $x_j\in\mathbb{C}$. In that case
 \begin{equation}\label{gu4b}
 \mathbf{{det}} \ X=x_0^2-\sum_{j=1}^3{x_j^2} \quad \mbox{and} \quad X^{-1}=\frac{1}{\mathbf{{det}} \ X}\left(x_0\sigma_0-\sum_{j=1}^3{x_j\sigma_j}\right).
 \end{equation}
In particular, if \ $X=\sigma_0-2(1+ik)\mathfrak{T}$, \ then
\begin{equation}\label{kkk2}
\mathbf{{det}} [\sigma_0-2(1+ik)\mathfrak{T}]=4(1+ik)^2\mathbf{{det} \ \mathfrak{T}}-4(1+ik)\gamma_0+1
\end{equation}
and
\begin{equation}\label{gu11}\fl
[\sigma_0-2(1+ik)\mathfrak{T}]^{-1}=\displaystyle{\frac{(1-2(1+ik)\gamma_0)\sigma_0+2(1+ik)\sum_{j=1}^3\gamma_j\sigma_j}{4(1+ik)^2\mathbf{{det} \ \mathfrak{T}}-4(1+ik)\gamma_0+1}},
\end{equation}
where $\gamma_j\in\mathbb{C}$ are determined uniquely by the decomposition
$$
\mathfrak{T}=\sum_{j=0}^3\gamma_j\sigma_j, \qquad \mbox{and} \qquad \mathbf{{det} \ \mathfrak{T}}=\gamma_0^2-\sum_{j=1}^3\gamma_j^2.
$$

Substituting (\ref{gu11}) into (\ref{red1}) and making elementary calculations we
obtain another representation of $S$-matrix of $A_{\mathfrak{T}}$
\begin{equation}\label{gu8}
{\textsf S}(k)=
\sigma_0+\displaystyle{4ik\frac{\mathfrak{T}-2(1+ik)\mathbf{{det} \ \mathfrak{T}}\sigma_0}{4(1+ik)^2\mathbf{{det} \ \mathfrak{T}}-4(1+ik)\gamma_0+1}}.
\end{equation}

The general formula (\ref{gu8}) can be simplified if we will consider separately the cases $\mathbf{{det} \ \mathfrak{T}}\not=0$
and $\mathbf{{det} \ \mathfrak{T}}=0.$
Denote $\theta_k=2(1+ik)$ and assume that $\mathbf{{det} \ \mathfrak{T}}\not=0$.
Then
\begin{equation}\label{gu32}
4(1+ik)^2\mathbf{{det} \ \mathfrak{T}}-4(1+ik)\gamma_0+1=\frac{(\theta_k-\theta_+)(\theta_k-\theta_-)}{\theta_-\theta_+},
\end{equation}
where
\begin{equation}\label{j5}\fl
\theta_+=\displaystyle{\frac{1}{\gamma_0+\sqrt{\sum_{j=1}^3\gamma_j^2}}}, \qquad \theta_-=\displaystyle{\frac{1}{\gamma_0-\sqrt{\sum_{j=1}^3\gamma_j^2}}}, \qquad  \mathbf{{det} \ \mathfrak{T}}=\frac{1}{\theta_-\theta_+}.
\end{equation}
Therefore, (\ref{gu8}) can be rewritten as
\begin{equation}\label{gu8c}
{\textsf S}(k)=\sigma_0+ 4ik\frac{\theta_-\theta_+\mathfrak{T}-\theta_k\sigma_0}{(\theta_k-\theta_+)(\theta_k-\theta_-)}.
\end{equation}
The decomposition of ${\textsf S}(k)$ with respect to the Pauli matrices
has the form
\begin{equation}\label{new3}
{\textsf S}(k)=\sum_{j=0}^3s_j(k)\sigma_j,
\end{equation}
where
\begin{equation}\label{au1}\fl
s_0(k)=1+4ik\frac{\theta_-\theta_+\gamma_0-\theta_k}{(\theta_k-\theta_+)(\theta_k-\theta_-)}, \quad s_j(k)=4ik\frac{\theta_-\theta_+\gamma_j}{(\theta_k-\theta_+)(\theta_k-\theta_-)}, \quad j\geq{1}.
\end{equation}

Let $\mathbf{{det} \ \mathfrak{T}}=0$. Then at least one of $\theta_\pm$ is equal to $\infty$ and (\ref{gu8c}) is reduced
to
\begin{equation}\label{gu8b}
{\textsf S}(k)=\sigma_0+\displaystyle{\frac{4ik}{1-2\theta_k\gamma_0}}\mathfrak{T}.
\end{equation}

\smallskip

Sometimes it is useful to calculate the $S$-matrix directly in terms of
 coefficients $a,b,c,d$ of the initial singular potential (\ref{gu14}). This means that we consider the particular case where
$A_{\mathfrak{T}}{\equiv}A_{\mathbf{T}}$ and  $\mathfrak{T}$ is defined by (\ref{gu3}). In that case the coefficients $\gamma_j$ of the decomposition
 $\mathfrak{T}=\sum_{j=0}^3\gamma_j\sigma_j$ have the form
\begin{equation}\label{gu9}
\begin{array}{lr}
\gamma_0=\displaystyle{\frac{1}{4\Xi}(\Xi-2(a+d))},  &  \gamma_1=\displaystyle{\frac{1}{4\Xi}(4+\mathbf{{det} \ T})}, \vspace{3mm} \\
\gamma_2=\displaystyle{\frac{-i}{2\Xi}(b-c)},  &  \gamma_3=\displaystyle{\frac{1}{2\Xi}(b+c)},
\end{array}
\end{equation}
where $\Xi=4-\mathbf{{det} \ T} + 2(a-d).$
Furthermore, the identities
\begin{equation}\label{gu10}
\mathbf{{det} \ \mathfrak{T}}=\displaystyle{-\frac{d}{2\Xi}}, \qquad \sum_{j=1}^3\gamma_j^2=\displaystyle{\frac{(4+\mathbf{{det} \ T})^2+16bc}{16\Xi^2}}
\end{equation}
 are deduced directly from (\ref{gu3}) and (\ref{gu9}).  (We remind that $\Xi$ is always non-zero due to our assumption  $-1\in\rho(A_{\mathbf{T}})$, see (\ref{gu4}).)
 Substituting the obtained relations into (\ref{gu8}) we
 obtain the expression of ${\textsf S}(\cdot)$ in terms of the coefficients $a,b,c,d$.
In particular,  if $d=0$, then $\mathbf{{det} \ \mathfrak{T}}=0$ and the expression (\ref{gu8}) is reduced to
\begin{equation}\label{gu88b}
{\textsf S}(k)=\sigma_0+\displaystyle{\frac{4ik\Xi}{2a(1+ik)-ik\Xi}}\mathfrak{T}.
\end{equation}

{\bf Example I.} \emph{$\delta$-potential with a complex coupling.} Let $a\in\mathbb{C}$ and $b=c=d=0$. Then (\ref{lesia11}) takes the form
$$
 -\frac{d^2}{dx^2}+a<\delta,\cdot>\delta(x), \qquad a\in\mathbb{C}
$$
and (\ref{lesia800}) determines the operators $A_{\mathbf{T}}\equiv{A_a}=\displaystyle{-\frac{d^2}{dx^2}}$ with domains of definition
$$
 \mathcal{D}(A_a)=\left\{f\in{{W_2^2}(\mathbb{R}\backslash\{0\})} \  \left|\right.  \begin{array}{l}
 f(0+)=f(0-) \ (\equiv{f(0)}) \vspace{2mm} \\
 f'(0+)-f'(0-)=af(0)\end{array} \right\}.
 $$
The matrix $\mathfrak{T}$ in (\ref{gu3}) and $\Xi$ are
$$
\mathfrak{T}=\frac{1}{4+2a}\left(\begin{array}{cc}
1 & 1 \\
1  & 1
\end{array}\right), \qquad  \Xi=4+2a.
$$

By virtue of (\ref{gu88b}),
$$
{\textsf S}(k)=\frac{1}{2k+ia}\left(\begin{array}{cc}
ia & -2k \\
-2k & ia
\end{array}\right).
$$

{\bf Example II.} \emph{Mixed complex $\delta$-potential.}
Let $b\in\mathbb{C}$ and $a=c=d=0$. Then (\ref{lesia11}) is reduced to
$$
 -\frac{d^2}{dx^2}+b<\delta',\cdot>\delta(x), \qquad b\in\mathbb{C}
$$
and domains of definition the corresponding operators $A_{\mathbf{T}}\equiv{A_b}=\displaystyle{-\frac{d^2}{dx^2}}$ take the form
$$
 \mathcal{D}(A_b)=\left\{f\in{{W_2^2}(\mathbb{R}\backslash\{0\})} \  \left|\right.  \begin{array}{l}
 f(0+)=f(0-) \ (\equiv{f(0)}) \vspace{2mm} \\
 (2+b)f'(0+)=(2-b)f'(0-) \end{array} \right\}.
 $$
The matrix $\mathfrak{T}$ and $\Xi$ are
$$
\mathfrak{T}=\frac{1}{8}\left(\begin{array}{cc}
2+b & 2-b \\
2+b  & 2-b
\end{array}\right), \qquad  \Xi=4.
$$

Using (\ref{gu88b}) again we obtain
$$
{\textsf S}(k)=\frac{-1}{2}\left(\begin{array}{cc}
b & 2-b \\
2+b & -b
\end{array}\right).
$$

{\bf Example III.} \emph{The case where the $S$-matrix is a constant on $\mathbb{C}$.}
The $S$-matrices of operators $A_b$ in Example II do not depend on $k$ and they are constants on $\mathbb{C}$.

Let $A_{\mathfrak{T}}$ be an operator defined by (\ref{sese2c}) and let
${\textsf S}_{\mathfrak{T}}(\cdot)$ be the corresponding $S$-matrix. An elementary analysis
of (\ref{gu8c}), (\ref{au1}), and (\ref{gu8b}) shows that  \emph{${\textsf S}_{\mathfrak{T}}(k)$ does not depend on
$k\in\mathbb{C}$ if and only if $\mathfrak{T}=0$, \  $\mathfrak{T}=\frac{1}{2}\sigma_0$, or}
$$
\mathfrak{T}=\frac{1}{4}\sigma_0+\sum_{j=1}^3\gamma_j\sigma_j, \quad  \mbox{where} \quad \sum_{j=1}^3\gamma_j^2=\frac{1}{16}.
$$
In these cases, respectively,
$$
{\textsf S}_{0}(k)=\sigma_0, \qquad  {\textsf S}_{\frac{1}{2}\sigma_0}(k)=-\sigma_0, \qquad {\textsf S}_{\mathfrak{T}}(k)=-4\sum_{j=1}^3\gamma_j\sigma_j.
$$

Assume now that the matrix $\mathfrak{T}$ can be expressed via ${\mathbf{T}}$ and hence,
$A_{\mathfrak{T}}\equiv{A_{\mathbf{T}}}$.  Using, (\ref{gu9}), (\ref{gu10}), and (\ref{gu88b}), we conclude
that the \emph{$S$-matrix ${\textsf S}_{\mathbf{T}}(k)$ of $A_{\mathbf{T}}$ is a constant on $\mathbb{C}$ if and only if
$a=d=0$.}

\subsection{The presentation of $S$-matrix in terms of transmission  and reflection  coefficients}
 The expression (\ref{red1}) of the $S$-matrix was obtained within the framework of the Lax-Phillips scattering theory and, certainly, it looks quite unusual.  Our aim now is to rewrite (\ref{red1}) in terms of transmission  and reflection  coefficients of the wave functions
\begin{equation}\label{deder1}
f_1=\left\{\begin{array}{ll}
{e^{-i\overline{{k}}x}}+R_{k}^{r}e^{i{k}x}, & x>0 \\
T_{k}^{r}e^{-i{k}x}, & x<0
\end{array}\right., \ \
f_2=\left\{\begin{array}{ll}
{T_{k}^l}e^{i{kx}}, & x>0 \\
e^{i\overline{{k}}x}+{R_{k}^l}e^{-i{k}x}, & x<0
\end{array}\right.
\end{equation}
where $k\in\mathbb{C}'=\mathbb{C}\setminus{i}\mathbb{R}=\{k\in\mathbb{C} : \textsf{Re}\ k\not=0 \}$.

It follows from (\ref{deder1}) that:
$$
\begin{array}{llll}
f_1(0+)=1+R^r_{k}, & f_1(0-)=T_{k}^r, &  f_1'(0+)=i(-\overline{{k}}+{k}R^r_{k}), & f_1'(0-)=-i{k}T_{k}^{k}, \vspace{2mm} \\
f_2(0+)=T_{k}^l,  & f_2(0-)=1+R^l_{k}, & f_2'(0+)=i{k}T_{k}^l, & f_2'(0-)=i(\overline{{k}}-{k}R^l_{k}).
\end{array}
$$
Substituting these values in (\ref{sese2c}) and solving the corresponding systems of linear equations, we get
$$
\begin{array}{ll}
\displaystyle{{\mathfrak t}_{11}=\frac{1}{\theta_k\Delta_{k}}[\Delta_{k}-(e^{i\alpha}-1)(R_{k}^l+e^{i\alpha})]}, \quad & \quad
\displaystyle{{\mathfrak t}_{12}=\frac{T_{k}^l}{\theta_k\Delta_{k}}(e^{i\alpha}-1)}, \vspace{4mm} \\
\displaystyle{{\mathfrak t}_{22}=\frac{1}{\theta_k\Delta_{k}}[\Delta_{k}-(e^{i\alpha}-1)(R_{k}^r+e^{i\alpha})]}, \quad & \quad
\displaystyle{{\mathfrak t}_{21}=\frac{T_{k}^r}{\theta_k\Delta_{k}}(e^{i\alpha}-1)},
\end{array}
$$
where $\theta_k=2(1+i{k})$, \ $\displaystyle{e^{i\alpha}=\frac{\overline{\theta_k}}{\theta_k}}$, \ $k\in\mathbb{C}_+'$, and
$$
\Delta_k=\left|\begin{array}{cc}
R_k^r+e^{i\alpha},  &  T_k^r \\
T_k^l,  & R_k^l+e^{i\alpha}
\end{array}\right|.
$$
Then
$$
\sigma_0-2(1+ik)\mathfrak{T}=\frac{e^{i\alpha}-1}{\Delta_k}\left(\begin{array}{cc}
R_k^l+e^{i\alpha} & -T_k^l  \vspace{4mm} \\
-T_k^r & R_k^r+e^{i\alpha}
\end{array}\right)
$$
and
$$
\det[\sigma_0-2(1+ik)\mathfrak{T}]=\frac{(e^{i\alpha}-1)^2}{\Delta_k}.
$$
Hence,
\begin{equation}\label{gu1}
[\sigma_0-2(1+ik)\mathfrak{T}]^{-1}=\frac{1}{e^{i\alpha}-1}\left(\begin{array}{cc}
R_k^r+e^{i\alpha} & T_k^l  \vspace{4mm} \\
T_k^r & R_k^l+e^{i\alpha}
\end{array}\right).
\end{equation}

Rewriting (\ref{red1}) as
$$
{\textsf S}(k)=\frac{1-ik}{1+ik}\sigma_0+\frac{2ik}{1+ik}[\delta_0-2(1+ik){\mathfrak T}]^{-1}, \qquad k\in\mathbb{C}_+',
$$
using (\ref{gu1}), and taking into account that
$$
\frac{2ik}{1+ik}\cdot\frac{1}{e^{i\alpha}-1}=-\frac{k}{\textsf{Re}\ k}, \quad
\frac{1-ik}{1+ik}-\frac{k}{\textsf{Re}\ k}e^{i\alpha}=-i\frac{\textsf{Im} \ k}{\textsf{Re} \ k}
$$
we obtain
\begin{equation}\label{rest2}
{\textsf S}(k)=-\frac{k}{\textsf{Re}\ k}\left(\begin{array}{cc}
R_k^r+\displaystyle{i\frac{\textsf{Im}\ k}{k}} & T_k^l  \vspace{4mm} \\
T_k^r & R_k^l+\displaystyle{i\frac{\textsf{Im}\ k}{k}}
\end{array}
\right).
\end{equation}

The expression (\ref{rest2}) coincides with the $S$-matrix of $A_{\mathfrak{T}}$
for all $k\in\mathbb{C}'$ such that
$$
\Delta_k=\left|\begin{array}{cc}
R_k^r+e^{i\alpha},  &  T_k^r \\
T_k^l,  & R_k^l+e^{i\alpha}
\end{array}\right|\not=0, \qquad e^{i\alpha}=\frac{1-i\overline{k}}{1+ik}.
$$

\setcounter{equation}{0}

\section{Similarity to self-adjoint operators}
 An operator $A$ acting in a Hilbert space $\mathfrak{H}$ is called \emph{similar} to a
 self-adjoint operator $H$  if there exists a bounded and boundedly invertible operator
 $Z$ such that
 \begin{equation}\label{bbbrrr1}
 A=Z^{-1}HZ.
\end{equation}
It is known (see, for example, \cite{KG}) that \emph{the similarity of $A$ to a self-adjoint operator means that $A$ is self-adjoint for a certain choice of inner product of the Hilbert space $\mathfrak{H}$, which is equivalent
to the initial inner product $(\cdot,\cdot)$}.

\smallskip

The following integral-resolvent criterion of similarity can be useful:
\begin{lemma}[\cite{NA}]\label{bibi}
 A closed densely defined operator $A$ acting in $\mathfrak{H}$
 is similar to a self-adjoint one if and only if the spectrum of $A$
 is real and there exists a constant $M$ such that
 \begin{equation}\label{bebe84}
 \begin{array}{l}
 \mathrm{sup}_{\varepsilon>0}\varepsilon\int_{-\infty}^{\infty}\|(A-zI)^{-1}g\|^2d\xi\leq{M}\|g\|^2, \vspace{2mm} \\
  \mathrm{sup}_{\varepsilon>0}\varepsilon\int_{-\infty}^{\infty}\|(A^*-zI)^{-1}g\|^2d\xi\leq{M}\|g\|^2,
\quad  \forall{g}\in\mathfrak{H},
 \end{array}
 \end{equation}
 where the integrals are taken along the line $z=\xi+i\varepsilon$ ($\varepsilon>0$
 is fixed)  of  $\mathbb{C}_+$.
\end{lemma}

In order to use Lemma \ref{bibi} we need an explicit form of the
resolvent $(A_{\mathfrak{T}}-zI)^{-1}$.

\begin{lemma}\label{l1}
 Let $A_{\mathfrak{T}}$ and $A_F$ be linear operators in $L_2(\mathbb{R})$ defined, respectively, by (\ref{sese2c}) and (\ref{new5}).
 Then, for all $g\in{L}_2(\mathbb{R})$ and for all $z=k^2$ ($k\in\mathbb{C}_+$) from the resolvent set of $A_{\mathfrak{T}}$
 $$
\|[(A_{\mathfrak{T}}-zI)^{-1}-(A_F-zI)^{-1}]g\|^2=\frac{1}{\textsf{Im}\
k}\left\|\frac{(\sigma_0+i\sigma_2)[\mathfrak{T}-\theta_k\mathbf{{det} \ \mathfrak{T}}\sigma_0]}{p_{{\mathfrak{T}}}(k)}Fg\right\|^2_{\mathbb{C}^2},
$$
where $\theta_k=2(1+ik)$, \ $\|\cdot\|_{\mathbb{C}^2}$ is the norm in $\mathbb{C}^2$, $Fg=\left(\begin{array}{c}
\int_{0}^{\infty}e^{iks}g(s)ds \vspace{2mm} \\
\int^{0}_{-\infty}e^{-iks}g(s)ds \end{array} \right)$
and
\begin{equation}\label{bbb16a}
p_{{\mathfrak{T}}}(k)=4(1+ik)^2\mathbf{{det} \ \mathfrak{T}}-4(1+ik)\gamma_0+1.
\end{equation}
\end{lemma}

\smallskip

 \emph{Proof.} Let us fix $k\in\mathbb{C}_+$ and consider the functions
\begin{equation}\label{sas2}\fl
 h_{1k}(x)=\left\{\begin{array}{cc}
 e^{ik{x}}, & x>0  \\
 e^{-ik{x}}, & x<0
 \end{array}\right.    \hspace{10mm}
 h_{2k}(x)=\left\{\begin{array}{cc}
 -e^{ik{x}} , & x>0  \\
 e^{-ik{x}}, & x<0
 \end{array}\right.
 \end{equation}
which belong $L_2(\mathbb{R})$ and form a basis of $\ker(A_{s}^*-zI)$, where $z=k^2\in\mathbb{C}\backslash\mathbb{R}_+$
and $A_{s}^*$ is the adjoint of the symmetric operator $A_s$ defined by (\ref{new9}).
Similarly to the proof of Lemma 4 in \cite{AlKuz}, we conclude that
\begin{equation}\label{new11}\fl
[(A_{\mathfrak{T}}-zI)^{-1}-(A_F-zI)^{-1}]g=c_{1k}h_{{1k}}+c_{2k}h_{{2k}}, \qquad  \forall{g}\in{L_2(\mathbb{R})},
\end{equation}
where  $c_{jk}$ are two parameters to be calculated. The latter relation allows one to express any function
$f\in\mathcal{D}(A_{\mathfrak{T}})$ as follows:
\begin{equation}\label{sas33}
f(x)=f_F(x)+c_{1k}h_{{1k}}(x)+c_{2k}h_{{2k}}(x),
\end{equation}
where $f_F=(A_F-zI)^{-1}g\in\mathcal{D}(A_F)$ and $f_F(0+)=f_F(0-)=0$ (in view of (\ref{new5})).

The functions $f$ in (\ref{sas33}) satisfy (\ref{sese2c}). Calculating the values of $f(0\pm), f'(0\pm)$ with the help
of (\ref{sas2}) and (\ref{sas33}), substituting them to (\ref{sese2c}) and making elementary transformations we get
\begin{equation}\label{sas89}
\left(\begin{array}{c}
c_{1k} \\
c_{2k} \end{array}\right)=(\sigma_0+i\sigma_2){\mathfrak{T}}(\sigma_0-\theta_k{\mathfrak{T}})^{-1}\left(\begin{array}{c}
f'_F(0+) \\
-f'_F(0-) \end{array}\right).
\end{equation}

Simple calculation with the use of (\ref{gu11}) and properties of Pauli matrices gives
$$
(\sigma_0+i\sigma_2){\mathfrak{T}}(\sigma_0-\theta_k{\mathfrak{T}})^{-1}=\frac{(\sigma_0+i\sigma_2)[\mathfrak{T}-\theta_k\mathbf{{det} \ \mathfrak{T}}\sigma_0]}{p_{{\mathfrak{T}}}(k)}.
$$
On the other hand, taking into account the explicit expression of $(A_F-zI)^{-1}$:
$$
(A_F-zI)^{-1}g=\left\{\begin{array}{l}
\frac{e^{ikx}}{k}\int_0^xg(s)\sin{ks}ds+\frac{\sin{kx}}{k}\int_x^\infty{e^{iks}g(s)}ds, \quad x>0; \vspace{3mm} \\
-\frac{e^{-ikx}}{k}\int^0_xg(s)\sin{ks}ds-\frac{\sin{kx}}{k}\int^x_{-\infty}{e^{-iks}g(s)}ds, \quad x<0
\end{array}\right.
$$
we obtain
$$
\left(\begin{array}{c}
f'_F(0+) \vspace{2mm} \\
-f'_F(0-) \end{array}\right)=\left(\begin{array}{c}
\int_{0}^{\infty}e^{iks}g(s)ds \vspace{2mm} \\
\int^{0}_{-\infty}e^{-iks}g(s)ds \end{array}\right).
$$
Thus, (\ref{sas89}) can be rewritten as
$$
\left(\begin{array}{c}
c_{1k} \\
c_{2k} \end{array}\right)=\frac{(\sigma_0+i\sigma_2)[\mathfrak{T}-\theta_k\mathbf{{det} \ \mathfrak{T}}\sigma_0]}{p_{{\mathfrak{T}}}(k)}Fg,  \quad \mbox{where} \quad
Fg=\left(\begin{array}{c}
\int_{0}^{\infty}e^{iks}g(s)ds \vspace{2mm} \\
\int^{0}_{-\infty}e^{-iks}g(s)ds \end{array} \right).
$$

The functions $h_{jk}$ in (\ref{sas2}) are orthogonal in $L_2(\mathbb{R})$ and $\|h_{jk}\|^2=\frac{1}{\textsf{Im}\
k}$. Hence, (\ref{new11}) gives
$$
\|[(A_{\mathfrak{T}}-zI)^{-1}-(A_F-zI)^{-1}]g\|^2=\frac{|c_{1k}|^2+|c_{2k}|^2}{\textsf{Im}\
k}=\frac{1}{\textsf{Im}\
k}\left\|\frac{(\sigma_0+i\sigma_2)[\mathfrak{T}-\theta_k\mathbf{{det} \ \mathfrak{T}}\sigma_0]}{p_{{\mathfrak{T}}}(k)}Fg\right\|^2_{\mathbb{C}^2}
$$
that completes the proof of Lemma \ref{l1}
\rule{2mm}{2mm}

\smallskip

\begin{theorem}\label{new4}
If all poles of the $S$-matrix ${\textsf S}(\cdot)$ of $A_{\mathfrak{T}}$ lie on the nonphysical sheet $\mathbb{C}_-$, then
$A_{\mathfrak{T}}$ is similar to a self-adjoint operator.
\end{theorem}
\emph{Proof.}
The operator $A_F$  defined by (\ref{new5}) is  self-adjoint. Hence, it satisfies (\ref{bebe84})
and the inequalities
 \begin{equation}\label{bebe84b}\fl
 \begin{array}{l}
 \mathrm{sup}_{\varepsilon>0}\varepsilon\int_{-\infty}^{\infty}\|[(A_{\mathfrak{T}}-zI)^{-1}-(A_F-zI)^{-1}]g\|^2d\xi\leq{M}\|g\|^2, \vspace{3mm}
 \\
  \mathrm{sup}_{\varepsilon>0}\varepsilon\int_{-\infty}^{\infty}\|[(A_{\mathfrak{T}}^*-zI)^{-1}-(A_F-zI)^{-1}]g\|^2d\xi\leq{M}\|g\|^2,
\quad  \forall{g}\in{L_2(\mathbb{R})},
 \end{array}
 \end{equation}
 give us the necessarily and sufficient condition for the similarity of $A_{\mathfrak{T}}$ to a self-adjoint operator.

Firstly we consider the auxiliary self-adjoint operator
$A_K$ defined by (\ref{new12}). Obviously,
the inequalities (\ref{bebe84b}) are true with $A_{\mathfrak{T}}={A_K}$.
Using Lemma \ref{l1}, and taking into account that
$$
A_K=A_{\frac{1}{2}\sigma_0}, \qquad  \mathbf{{det} \ \frac{1}{2}\sigma_0}=\frac{1}{4}, \qquad
 p_{\frac{1}{2}\sigma_0}(k)=-k^2
 $$  we get
 $$
 \|[(A_K-zI)^{-1}-(A_F-zI)^{-1}]g\|^2=\frac{1}{\textsf{Im}\
k}\left\|\frac{(\sigma_0+i\sigma_2)}{2k}Fg\right\|^2_{\mathbb{C}^2}.
 $$
 Therefore, in view of (\ref{bebe84b}),
\begin{equation}\label{des1}
\int_{-\infty}^{\infty}\frac{\varepsilon}{\textsf{Im}\
k}\left\|\frac{(\sigma_0+i\sigma_2)}{k}Fg\right\|^2_{\mathbb{C}^2}d\xi \ {\leq}  \ {M}\|g\|^2.
\end{equation}
We note that the integral in (\ref{des1}) is taken along the line $z=k^2=\xi+i\varepsilon$ ($\varepsilon>0$
 is fixed) of upper half-plane $\mathbb{C}_+$. This means that
 $$
 \varepsilon=2(\textsf{Re} \ k)(\textsf{Im} \ k)>0, \qquad \xi=(\textsf{Re} \ k)^2-(\textsf{Im} \ k)^2.
 $$
Therefore, the variable $k$ belongs to $\mathbb{C}_{++}=\{k\in\mathbb{C}_+ : \textsf{Re} \ k>0\}$ when $k^2=\xi+i\varepsilon$.

Let $A_{\mathfrak{T}}$ be an operator defined by (\ref{sese2c}). Assume that the $S$-matrix of
$A_{\mathfrak{T}}$  has poles on the nonphysical sheet $\mathbb{C}_-$ only.
Then, taking into account (\ref{gu8}) and (\ref{au1}), we conclude that the roots of
$p_{{\mathfrak{T}}}(k)$ belong to $\mathbb{C}_-$. Hence, the entries of the matrix
$$
 \Psi(k)=\frac{k}{p_{\mathfrak{T}}(k)}[\mathfrak{T}-\theta_k\mathbf{{det} \ \mathfrak{T}}\sigma_0]
$$
are uniformly bounded when $k$ runs $\mathbb{C}_{++}$.
Taking in mind this fact, Lemma \ref{l1} and (\ref{des1}) we obtain
$$
 \varepsilon\int_{-\infty}^{\infty}\|[(A_{\mathfrak{T}}-zI)^{-1}-(A_F-zI)^{-1}]g\|^2d\xi=
 \int_{-\infty}^{\infty}\frac{\varepsilon}{\textsf{Im}\
k}\left\|\frac{(\sigma_0+i\sigma_2)}{k}\Psi(k)Fg\right\|^2_{\mathbb{C}^2}d\xi\leq
$$
$$
M_1\int_{-\infty}^{\infty}\frac{\varepsilon}{\textsf{Im}\
k}\left\|\frac{(\sigma_0+i\sigma_2)}{k}Fg\right\|^2_{\mathbb{C}^2}d\xi<MM_1\|g\|^2
$$
that establish the first inequality in (\ref{bebe84b}).

The second inequality can be justified in a similar manner.
Indeed, it is easy to check that the domain of definition $\mathcal{D}(A_{\mathfrak{T}}^*)$ has the form
(\ref{sese2c}) with ${\mathfrak{T}}^*$ (instead of $\mathfrak{T}$). Thus,
\begin{equation}\label{new17}
A_{\mathfrak{T}}^*=A_{\mathfrak{T}^*}.
\end{equation}

Let ${\textsf S}_{\mathfrak{T}}(\cdot)$ and ${\textsf S}_{\mathfrak{T}^*}(\cdot)$ be the $S$-matrix of operators
$A_{\mathfrak{T}}$ and $A_{\mathfrak{T}^*}$, respectively. It follows from (\ref{red1}) that
\begin{equation}\label{new7}
{\textsf S}_{\mathfrak{T}^*}(-\overline{k})={\textsf S}_{\mathfrak{T}}^*(k),  \qquad k\in\mathbb{C}.
\end{equation}
Therefore, the $S$-matrix of $A_{\mathfrak{T}^*}$ also has poles within $\mathbb{C}_-$.
This allows one to establish the second relation in (\ref{bebe84b}) by repeating the previous arguments
with the use of modified matrix
$$
\Psi(k)=\frac{k}{p_{\mathfrak{T}^*}(k)}[\mathfrak{T}^*-\theta_k\mathbf{{det} \ \mathfrak{T}^*}\sigma_0].
$$

In view of Lemma \ref{bibi} and inequalities (\ref{bebe84b}) the operator
$A_{\mathfrak{T}}$ is similar to a self-adjoint one.
Theorem \ref{new4} is proved.
\rule{2mm}{2mm}

\begin{corollary}
Let the $S$-matrix of $A_{\mathfrak{T}}$ be a constant on $\mathbb{C}$ (see Example III). Then $A_{\mathfrak{T}}$ is similar
to a self-adjoint operator.
\end{corollary}

\setcounter{equation}{0}

\section{Metric operators}
Unfortunately, the proof of Theorem \ref{new4} does not contain `a recipe' of construction of an appropriative metric operator which
guarantees the self-adjointness of $A_{\mathfrak{T}}$. We just state that such an operator exists.
Various approaches to the explicit determination of metric operator with the use of formal perturbative methods as well as
mathematically rigid constructions can be found in \cite{MET}.

In this section we are aiming to find an explicit expression for metric operators in the case where
the $S$-matrix  ${\textsf S}(\cdot)$ of $A_{\mathfrak{T}}$ \emph{has simple non-zero imaginary poles.}

\smallskip

Assume that $Q$ is a self-adjoint operator in $L_2(\mathbb{R})$. Then $e^{\chi{Q}}$, $(\chi\in\mathbb{R})$
is a positive self-adjoint operator in $L_2(\mathbb{R})$. If there exists a metric operator $e^{\chi{Q}}$ such that
\begin{equation}\label{new12c}
e^{\chi{Q}}A_{\mathfrak{T}}=A_{\mathfrak{T}}^*e^{\chi{Q}},
\end{equation}
then $A_{\mathfrak{T}}$ turns out to be
self-adjoint with respect to the new inner product $\|\cdot\|_{new}^2=(e^{\chi{Q}}\cdot, \cdot)=\|e^{\chi{Q}/2}\cdot\|$
of $L_2(\mathbb{R})$.

Using (\ref{new17}) we rewrite (\ref{new12c}) in the equivalent form
\begin{equation}\label{new12b}
e^{\chi{Q}}A_{\mathfrak{T}}=A_{\mathfrak{T}^*}e^{\chi{Q}}
\end{equation}
and we will seek the operator $Q$ in (\ref{new12b}) as:
\begin{equation}\label{new13}
Q_{\vec{\alpha}}=\alpha_1\mathcal{P}+\alpha_2{i}\mathcal{P}\mathcal{R}+\alpha_3\mathcal{R}, \quad
\vec{\alpha}=(\alpha_1, \alpha_2, \alpha_3)\in\mathbb{S}^2,
\end{equation}
where $\mathbb{S}^2=\{\vec{\alpha}\in\mathbb{R}^3 \ : \ \sum_{j=1}^3\alpha_j^2=1\}$  and
\begin{equation}\label{new144}
\mathcal{P}f(x)=f(-x), \quad \mathcal{R}f(x)=(\textsf{sgn}\ x)f(x), \quad \forall{f}\in{L_2(\mathbb{R})}
\end{equation}
are self-adjoint operators in $L_2(\mathbb{R})$.

The operators $Q_{\vec{\alpha}}$ are self-adjoint in $L_2(\mathbb{R})$ and $Q_{\vec{\alpha}}^2=I$ \cite{NEW}.
Therefore,
\begin{equation}\label{new15}
e^{\chi{Q_{\vec{\alpha}}}}=(\cosh\chi){I}+(\sinh{\chi})Q_{\vec{\alpha}}, \qquad  \chi\in\mathbb{R}.
\end{equation}

It follows from (\ref{new13}) -- (\ref{new15}) that $e^{\chi{Q_{\vec{\alpha}}}}$ commutes
with the operator
$$
A_s^*=-\frac{d^2}{dx^2}, \qquad \mathcal{D}(A_{s}^*)={W}_2^2(\mathbb{R}\backslash\{0\}).
$$

Since $A_{\mathfrak{T}}$ and $A_{\mathfrak{T}^*}$ are restrictions of $A_s^*$, respectively, onto
$\mathcal{D}(A_{\mathfrak{T}})$ and $\mathcal{D}(A_{\mathfrak{T}^*})$
the relation (\ref{new12b}) holds if and only if the operator $e^{\chi{Q_{\vec{\alpha}}}}$ maps
$\mathcal{D}(A_{\mathfrak{T}})$ into  $\mathcal{D}(A_{\mathfrak{T}^*})$,
Taking (\ref{sese2c}) into account we conclude that the relation
$e^{\chi{Q_{\vec{\alpha}}}} : \mathcal{D}(A_{\mathfrak{T}}) \to \mathcal{D}(A_{\mathfrak{T}^*})$ is
equivalent to the following implication
\begin{equation}\label{lesia701}\fl
\mbox{if} \ \hspace{5mm} \mathfrak{T}\Gamma_0f=\Gamma_1f,
\hspace{5mm} \ \mbox{then} \ \hspace{5mm}
{\mathfrak{T}}^*\Gamma_0e^{\chi{Q_{\vec{\alpha}}}}f=\Gamma_1e^{\chi{Q_{\vec{\alpha}}}}f, \qquad \forall{f}\in{\mathcal{D}(A_{\mathfrak{T}})},
\end{equation}
where $\Gamma_0f=\left(\begin{array}{c}
f(0+)+f'(0+) \\
f(0-)-f'(0-)
\end{array}\right)$ \ and \ $\Gamma_1f=\displaystyle{\frac{1}{2}}\left(\begin{array}{c}
f(0+) \\
f(0-)
\end{array}\right)$.

It is easily to check, using  the  definition (\ref{new144}) of operators $\mathcal{P}$ and $\mathcal{R}$, that
$$
\Gamma_k\mathcal{P}f=\sigma_1\Gamma_kf, \quad \Gamma_k\mathcal{R}f=\sigma_3\Gamma_kf, \quad \Gamma_k{i}\mathcal{P}\mathcal{R}f=i\sigma_1\sigma_3\Gamma_kf=\sigma_2\Gamma_kf, \quad \forall{f}\in{{W}_2^2(\mathbb{R}\backslash\{0\})}.
$$
Therefore,
$\Gamma_ke^{\chi{Q_{\vec{\alpha}}}}f=(\cosh\chi\sigma_0+\sinh{\chi}\sigma_{\vec{\alpha}})\Gamma_kf, \ k=0, 1$,
where $\sigma_{\vec{\alpha}}=\sum_{j=1}^3\alpha_j\sigma_j$ and implication (\ref{lesia701}) is equivalent to equation
\begin{equation}\label{new21}
{\mathfrak{T}}^*(\cosh\chi\sigma_0+\sinh{\chi}\sigma_{\vec{\alpha}})=(\cosh\chi\sigma_0+\sinh{\chi}\sigma_{\vec{\alpha}}){\mathfrak{T}}
\end{equation}
with respect to unknown $\chi\in\mathbb{R}$ and $\vec{\alpha}\in\mathbb{S}^2$.

Assume that all poles of $S$-matrix  ${\textsf S}(\cdot)$ of a \emph{non-self-adjoint operator} $A_{\mathfrak{T}}$ are simple and they
belong to $\mathbb{R}\setminus\{0\}$. In view of (\ref{j5}) -- (\ref{gu8b}),
the case of  \emph{two different simple non-zero imaginary poles} of ${\textsf S}(\cdot)$ is characterized
by the conditions
\begin{equation}\label{new22}
\mathbf{{det} \ \mathfrak{T}}\not=0, \qquad \gamma_0\in\mathbb{R}, \qquad  \sqrt{\sum_{j=1}^3\gamma_j^2}\in\mathbb{R}\setminus\{0\},
\end{equation}
where (as usual) $\mathfrak{T}=\sum_{j=0}^3\gamma_j\sigma_j$.
Similarly the case where $S$-matrix of a non-self-adjoint operator $A_{\mathfrak{T}}$
has \emph{one simple non-zero imaginary pole} corresponds to the relations
\begin{equation}\label{new22b}
\mathbf{{det} \ \mathfrak{T}}=0, \qquad \gamma_0\in\mathbb{R}, \qquad  \sqrt{\sum_{j=1}^3\gamma_j^2}\in\mathbb{R}\setminus\{0\}.
\end{equation}

The condition $\gamma_0\in\mathbb{R}$ in both cases (\ref{new22}) and (\ref{new22b}) allows to rewrite
the equation (\ref{new21}) as follows:
\begin{equation}\label{new24}\fl
\sum_{j=1}^3({\textsf{Im}\ \gamma_j})\sigma_j=\tanh{\chi}\left(\left|\begin{array}{ccc}
\sigma_1 & \sigma_2 & \sigma_3  \\
\textsf{Re}\ \gamma_1 & \textsf{Re}\ \gamma_2 & \textsf{Re}\ \gamma_3 \\
\alpha_1 & \alpha_2 & \alpha_3
\end{array} \right| - \sum_{j=1}^3({\textsf{Im}\ \gamma_j})\alpha_j\sigma_0\right),
\end{equation}
where the formal determinant
$$
\left|\begin{array}{ccc}
\sigma_1 & \sigma_2 & \sigma_3  \\
\textsf{Re}\ \gamma_1 & \textsf{Re}\ \gamma_2 & \textsf{Re}\ \gamma_3 \\
\alpha_1 & \alpha_2 & \alpha_3
\end{array} \right|:=\left|\begin{array}{cc}
\textsf{Re}\ \gamma_2 & \textsf{Re}\ \gamma_3 \\
\alpha_2 & \alpha_3
\end{array}
\right|\sigma_1-\left|\begin{array}{cc}
\textsf{Re}\ \gamma_1 & \textsf{Re}\ \gamma_3 \\
\alpha_1 & \alpha_3
\end{array}
\right|\sigma_2+\left|\begin{array}{cc}
\textsf{Re}\ \gamma_1 & \textsf{Re}\ \gamma_2 \\
\alpha_1 & \alpha_2
\end{array}
\right|\sigma_3
$$
is the `cross product' of vectors ${\textsf{Re}\ \vec{\gamma}}=(\textsf{Re}\ \gamma_1, \textsf{Re}\ \gamma_2, \textsf{Re}\ \gamma_3)$ and $\vec{\alpha}$ which is associated with the Pauli matrices $\sigma_1, \sigma_2, \sigma_3$ (instead of the standard basis vectors $\mathbf{i}, \mathbf{j}, \mathbf{k}$ of the Euclidean space $\mathbb{R}^3$).

We remark that the vectors
$$
{\textsf{Re}\ \vec{\gamma}}=(\textsf{Re}\ \gamma_1, \textsf{Re}\ \gamma_2, \textsf{Re}\ \gamma_3), \qquad
{\textsf{Im}\ \vec{\gamma}}=(\textsf{Im}\ \gamma_1, \textsf{Im}\ \gamma_2, \textsf{Im}\ \gamma_3)
$$
in (\ref{new24}) cannot be zero. Indeed, if  ${\textsf{Re}\ \vec{\gamma}}=\vec{0}$, then $\sqrt{\sum_{j=1}^3\gamma_j^2}=\sqrt{-\sum_{j=1}^3|\gamma_j|^2}\in{i}\mathbb{R}\setminus\{0\}$ that contradicts to the third relation in (\ref{new22}), (\ref{new22b}). Similarly, if ${\textsf{Im}\ \vec{\gamma}}=\vec{0}$, then the second relation in (\ref{new22}), (\ref{new22b})
implies that $A_{\mathfrak{T}}$ is a self-adjoint operator that is impossible.

It follows from the third relation in (\ref{new22}), (\ref{new22b}) that
$$
\sum_{j=1}^3\gamma_j^2=\sum_{j=1}^3(\textsf{Re}\ \gamma_j)^2-\sum_{j=1}^3(\textsf{Im}\ \gamma_j)^2+ 2i\sum_{j=1}^3(\textsf{Re}\ \gamma_j)(\textsf{Im}\ \gamma_j)>0.
$$
Hence,
\begin{equation}\label{new24a}
\sum_{j=1}^3(\textsf{Re}\ \gamma_j)^2>\sum_{j=1}^3(\textsf{Im}\ \gamma_j)^2, \qquad \sum_{j=1}^3(\textsf{Re}\ \gamma_j)(\textsf{Im}\ \gamma_j)=0.
\end{equation}
This means that the vectors ${\textsf{Re}\ \vec{\gamma}}$ and ${\textsf{Im}\ \vec{\gamma}}$ are orthogonal in $\mathbb{R}^3$.

\smallskip

Let us fix the vector $\vec{\alpha}\in\mathbb{S}^2$ in such a way that $\vec{\alpha}$ is orthogonal to ${\textsf{Re}\ \vec{\gamma}}$ and ${\textsf{Im}\ \vec{\gamma}}$. Then the standard cross product
${\textsf{Re}\ \vec{\gamma}}\times\vec{\alpha}=\left|\begin{array}{ccc}
\mathbf{i} & \mathbf{j} & \mathbf{k}  \\
\textsf{Re}\ \gamma_1 & \textsf{Re}\ \gamma_2 & \textsf{Re}\ \gamma_3 \\
\alpha_1 & \alpha_2 & \alpha_3
\end{array} \right|$ is collinear to ${\textsf{Im}\ \vec{\gamma}}$.
Precisely, there exists $\kappa\in\mathbb{R}$ such that
\begin{equation}\label{new28}
{\textsf{Im}\ \vec{\gamma}}=\kappa{\textsf{Re}\ \vec{\gamma}}\times\vec{\alpha}.
\end{equation}
Calculating the norms of vectors ${\textsf{Im}\ \vec{\gamma}}$ and ${\textsf{Re}\ \vec{\gamma}}\times\vec{\alpha}$ in (\ref{new28})
and taking into account (\ref{new24a}), we obtain
$$
|k|^2=\frac{\|{\textsf{Im}\ \vec{\gamma}}\|^2}{\|{\textsf{Re}\ \vec{\gamma}\|^2}\|\vec{\alpha}\|^2}=\frac{\|{\textsf{Im}\ \vec{\gamma}}\|^2}{\|{\textsf{Re}\ \vec{\gamma}\|^2}}=\displaystyle{\frac{\sum_{j=1}^3(\textsf{Im}\ \gamma_j)^2}{\sum_{j=1}^3(\textsf{Re}\ \gamma_j)^2}<1}.
$$

On the other hand, since $\vec{\alpha}\perp{\textsf{Im}\ \vec{\gamma}}$, the equation (\ref{new24}) takes the form
\begin{equation}\label{new24b}
\sum_{j=1}^3({\textsf{Im}\ \gamma_j})\sigma_j=\tanh{\chi}\left|\begin{array}{ccc}
\sigma_1 & \sigma_2 & \sigma_3  \\
\textsf{Re}\ \gamma_1 & \textsf{Re}\ \gamma_2 & \textsf{Re}\ \gamma_3 \\
\alpha_1 & \alpha_2 & \alpha_3
\end{array} \right|.
\end{equation}
Obviously, (\ref{new24b}) has a solution $\chi\in\mathbb{R}$ such that $\tanh{\chi}=k$.
Summing the results above, we prove
\begin{theorem}\label{new25b}
If the $S$-matrix ${\textsf S}(\cdot)$ of a non-self-adjoint operator $A_{\mathfrak{T}}$ has simple non-zero imaginary poles, then $A_{\mathfrak{T}}$ turns out to be self-adjoint with respect to new inner product $\|\cdot\|_{new}^2=(e^{\chi{Q_{\vec{\alpha}}}}\cdot, \cdot)$, where
$\vec{\alpha}\in\mathbb{S}^2$ is orthogonal to the vectors ${\textsf{Re}\ \vec{\gamma}}$, ${\textsf{Im}\ \vec{\gamma}}$ and
$\chi$ is defined by the relation $\tanh{\chi}=\kappa$, where $\kappa$ is the coefficient of collinearity in $(\ref{new28})$.
\end{theorem}

It looks natural that the parameter $\chi$ in Theorem \ref{new25b}
correlates to the distance between imaginary poles of ${\textsf S}(\cdot)$.

\begin{corollary}\label{new29}
Let $k_{\pm}$ be two imaginary poles of the $S$-matrix of $A_{\mathfrak{T}}$.
Then the parameter $\chi$ of the corresponding metric operator $e^{\chi{Q_{\vec{\alpha}}}}$ can be determined by the relation
\begin{equation}\label{new31}
\cosh\chi=\frac{\|\textsf{Re}\ \vec{\gamma}\|}{|(k_--k_+)\mathbf{{det} \ \mathfrak{T}}|}.
\end{equation}
\end{corollary}
\emph{Proof.} If $k_{\pm}$ are poles of ${\textsf S}(\cdot)$, then the quantities $\theta_{\pm}$ in (\ref{j5}) are
expressed as $\theta_{\pm}=2(1+i{k}_{\pm})$.
Denote $\xi=\sqrt{\sum_{j=1}^3\gamma_j^2}$. Then
$$
\xi=\frac{1}{2\theta_+}-\frac{1}{2\theta_-}=i(k_--k_+)\mathbf{{det} \ \mathfrak{T}}.
$$

Taking into account that $\xi^2=\sum_{j=1}^3(\textsf{Re}\ \gamma_j)^2-(\textsf{Im}\ \gamma_j)^2$, we obtain
 $$
 \left\|\frac{\textsf{Re}\ \vec{\gamma}}{\xi}\right\|^2-\left\|\frac{\textsf{Im}\ \vec{\gamma}}{\xi}\right\|^2=1.
 $$
 Therefore, there exists $\omega\geq0$ such that $\cosh\omega=\left\|\frac{\textsf{Re}\ \vec{\gamma}}{\xi}\right\|$ and
 $\sinh\omega=\left\|\frac{\textsf{Im}\ \vec{\gamma}}{\xi}\right\|$.

It follows from (\ref{new28}) and Theorem \ref{new25b} that
$$
|\tanh\chi|=|k|=\frac{\|\textsf{Im}\ \vec{\gamma}\|}{\|\textsf{Re}\ \vec{\gamma}\|}=\frac{\sinh\omega}{\cosh\omega}=\tanh\omega.
$$
Without loss of generality\footnote{by choosing an appropriative direction of $\vec{\alpha}$ in (\ref{new28})} we can suppose that $k\geq{0}$ in (\ref{new28}).
Then $\chi=\omega$ and $\cosh\chi$ is determined by (\ref{new31}). \rule{2mm}{2mm}

\setcounter{equation}{0}

\section{Spectral singularities}
If $A_{\mathfrak{T}}$ is a self-adjoint operator in $L_2(\mathbb{R})$ or $A_{\mathfrak{T}}$ is similar to a
self-adjoint one, then the entries of the $S$-matrix ${\textsf S}(k)$
are uniformly bounded when $k$ runs $\mathbb{R}$. Since
the existence of spectral singularity
$z=k^2_0$ of $A_{\mathfrak{T}}$ should mean that $A_{\mathfrak{T}}$ cannot be  similar to a self-adjoint operator,
it is natural to suppose that ${\textsf S}(k)$ cannot be uniformly bounded in a neighborhood
of $k_0\in\mathbb{R}$.  This leads to the following

\begin{definition}\label{def12}
A nonnegative number $z=k^2_0$ is called the spectral singularity of $A_{\mathfrak{T}}$  if
$k_0\in\mathbb{R}$ is a pole of the $S$-matrix ${\textsf S}(\cdot)$ of $A_{\mathfrak{T}}$. The operator $A_{\mathfrak{T}}$ has
spectral singularity at infinity if $k_0=\infty$ is a pole of ${\textsf S}(\cdot)$.
\end{definition}

It is known (see, e.g. \cite{AlKuz}) that the continuous spectrum of operators $A_{\mathfrak{T}}$ defined by
(\ref{sese2c}) coincides with $[0,\infty)$ and there are no eigenvalues of $A_{\mathfrak{T}}$ embedded in continuous spectrum.
Therefore, spectral singularities of $A_{\mathfrak{T}}$ may appear on the continuous spectrum only and
(possible) existence of a spectral singularity $z$  does not mean that $z$ is an eigenvalue $A_{\mathfrak{T}}$.

\begin{proposition}\label{new6}
The operators $A_{\mathfrak{T}}$ and $A_{\mathfrak{T}}^*$  have the same set of spectral singularities.
\end{proposition}
\emph{Proof.} Follows immediately from the relation $A_{\mathfrak{T}}^*=A_{\mathfrak{T}^*}$ and (\ref{new7}).
\rule{2mm}{2mm}

\smallskip

The existence of spectral singularity of $A_{\mathfrak{T}}$ can be easily described via the roots of the polynomial
$p_{{\mathfrak{T}}}(k)$ defined by (\ref{bbb16a}).

\begin{proposition}\label{neew2}
Assume that ${\mathfrak{T}}\not=\frac{1}{2}\sigma_0$. A point $z=k^2_0$ is a spectral singularity of $A_{\mathfrak{T}}$ if and only if the polynomial (\ref{bbb16a}) has:
\begin{enumerate}
\item  a root $k_0\in\mathbb{R}$ for the case of nonzero spectral singularity $z\not=0$;
\item a root $k_0=0$ of multiplicity 2 for the case of zero spectral singularity $z=0$;
\item no roots for the case of spectral singularity at $z=\infty$.
\end{enumerate}
\end{proposition}
\emph{Proof.}
Let $z=k_0^2\not=0$ be a spectral singularity of $A_{\mathfrak{T}}$. Then $k_0\in\mathbb{R}\setminus\{0\}$ is a pole of ${\textsf S}(k)$.
Assume firstly that $\mathbf{{det} \ \mathfrak{T}}\not=0$. Then ${\textsf S}(\cdot)$ is determined by
(\ref{gu8c}), where $\theta_-\theta_+\not=0$ and $\theta_-\theta_+\not\not=\infty$ due to the third relation in (\ref{j5}).
The existence of pole $k_0$ of ${\textsf S}(\cdot)$ means that $\theta_{k_0}=2(1+ik_0)$ coincides
with $\theta_-$ or with $\theta_+$. Then the point $k_0$ is a root of $p_{{\mathfrak{T}}}(k)$
due to (\ref{gu32}).  Conversely, if $k_0$ is a root of $p_{{\mathfrak{T}}}(k)$, then $k_0$ is a pole of ${\textsf S}(k)$
(this implication follows from (\ref{gu32}) -- (\ref{au1})).

Assume now that $\mathbf{{det} \ \mathfrak{T}}=0$. Then
${\textsf S}(\cdot)$ is determined by
(\ref{gu8b}). The pole $k_0$ of ${\textsf S}(\cdot)$ is possible where $-2\theta_{k_0}\gamma_0+1=-4(1+ik_0)\gamma_0+1=0$. Thus,
$k_0$ is a root of $p_{{\mathfrak{T}}}(k)$. Conversely statement is evident. Implication (i) is proved.

\smallskip

 Let $z=k_0^2=0$ be a spectral singularity of $A_{\mathfrak{T}}$.  Then $k_0=0$ is a pole of ${\textsf S}(k)$.
Let $\mathbf{{det} \ \mathfrak{T}}\not=0$. Then the $S$-matrix is determined by
(\ref{gu8c}) and simple analysis of (\ref{gu8c}) shows that ${\textsf S}(k)$ has a pole at $k_0=0$
in the case $\theta_-=\theta_+=\theta_0=2$ only.  By (\ref{gu32}), $k_0=0$ is a root of $p_{{\mathfrak{T}}}(k)$ of multiplicity
$2$. Conversely, let $k_0=0$ be a root of $p_{\mathfrak{T}}(k)$ of multiplicity $2$. Then $\theta_-=\theta_+=\theta_0=2$. Using
(\ref{gu8c}) again we deduce that ${\textsf S}(k)$ has a pole at $k_0=0$. (The case $\gamma_0=\frac{1}{2}$ and  $\gamma_1=\gamma_2=\gamma_3=0$ is not considered here because,
 $\mathfrak{T}\not=\frac{1}{2}\sigma_0$ by the assumption of Proposition \ref{neew2}.)

Assume now that $\mathbf{{det} \ \mathfrak{T}}=0$. Then
${\textsf S}(\cdot)$ is determined by (\ref{gu8b}) and this expression does not have a pole at $k_0=0$.
On the other hand, $k_0=0$ cannot be a root of $p_{{\mathfrak{T}}}(k)$ of multiplicity
$2$ when $\mathbf{{det} \ \mathfrak{T}}=0$. Implication (ii) is proved.

\smallskip

To prove (iii) it suffices to note that ${\textsf S}(k)$ will tend to infinity when $k\to\infty$ only in the case
where $p_{{\mathfrak{T}}}(k)$ does not depend on $k$. This means that
$p_{{\mathfrak{T}}}(k)$ has no roots in $\mathbb{C}$.
Proposition \ref{neew2} is proved.
\rule{2mm}{2mm}

\smallskip

The `exceptional' operator $A_{\frac{1}{2}\sigma_0}$ in Proposition \ref{neew2}
coincides with the Krein extension of the symmetric operator $A_s$ (see Remark \ref{rrr4}).

In the particular case where  $A_{\mathfrak{T}}=A_{\mathbf{T}}$, spectral singularities
are described via the roots of the polynomial
\begin{equation}\label{bbb16b}
 p_{{\mathbf{T}}}(k)=2dk^2+i(\mathbf{{det} \ T}-4)k+2a.
\end{equation}
\begin{corollary}[\cite{GROD}]\label{neeww22}
A point $z=k^2_0$ is a spectral singularity of $A_{\mathbf{T}}$ if and only if the polynomial (\ref{bbb16b}) has:
\begin{enumerate}
\item  a real root $k_0\in\mathbb{R}$ for the case of nonzero spectral singularity $z\not=0$;
\item  the zero root $k_0=0$ of multiplicity 2 for the case of spectral singularity at $z=0$;
\item no roots for the case of spectral singularity at $z=\infty$.
\end{enumerate}
\end{corollary}
\emph{Proof.} First of all we note that the domain of definition of $A_K(=A_{\frac{1}{2}\sigma_0})$
cannot be presented in the form (\ref{lesia800}). Thus $A_{\frac{1}{2}\sigma_0}$ cannot be realized as $A_{\mathbf{T}}$.
Taking the expressions of $\gamma_0$ and $\mathbf{{det} \ \mathfrak{T}}$ given by (\ref{gu9}) and (\ref{gu10}) into account,
 we get
 $p_{{\mathfrak{T}}}(k)=\frac{1}{\Xi}p_{{\mathbf{T}}}(k).$
This relation and Proposition \ref{neew2} complete the proof. \rule{2mm}{2mm}

\begin{proposition}\label{gg1}
If $A_{\mathfrak{T}}$ has a spectral singularity, then $A_{\mathfrak{T}}$  can not be similar to a self-adjoint operator.
\end{proposition}

\emph{Proof.}
The resolvent of an arbitrary self-adjoint operator $H$ satisfies the inequality $\|(H-zI)^{-1}\|\leq\frac{1}{|\textsf{Im}\
z|}$
for all $z\in\mathbb{C}\setminus\mathbb{R}$. If $A$ is similar to a self-adjoint operator (i.e., (\ref{bbbrrr1}) holds), then the inequality
above takes the form
\begin{equation}\label{bbb6}
\|(A-zI)^{-1}\|\leq\frac{C}{|\textsf{Im}\ z|}, \qquad C=\|Z^{-1}\|\|Z\|, \quad z\in\mathbb{C}\setminus\mathbb{R}.
\end{equation}

Let $A_{\mathfrak{T}}$ be similar to a self-adjoint operator. Since $A_F$ is self-adjoint, the relation  (\ref{bbb6})
holds for $A_{\mathfrak{T}}$ and for $A_F$. Therefore,
 \begin{equation}\label{sas40}
 \|[(A_{\mathfrak{T}}-zI)^{-1}-(A_F-zI)^{-1}]g\|^2\leq\frac{M}{(\textsf{Im}\ z)^2}\|g\|^2,
 \end{equation}
 where $M$ is a constant independent of $g\in{L_2(\mathbb{R})}$ and $z\in{\mathbb{C}\setminus\mathbb{R}}$.

Let us consider a particular case of (\ref{sas40}) with
$z=k^2 \ (k\in\mathbb{C}_{+})$ and
 $g=g_\pm$, where
 $$
 g_+(x)=\left\{\begin{array}{cc}
 e^{-i\overline{k}{x}}, & x>0;  \\
 0, & x<0
 \end{array}\right.    \hspace{10mm}
 g_-(x)=\left\{\begin{array}{cc}
 0 , & x>0  \\
 e^{i\overline{k}{x}}, & x<0
 \end{array}\right.
 $$
 Taking into account that
 \begin{equation}\label{bebe36}
 \|g_{\pm}\|^2=\frac{1}{2\textsf{Im}\ k}, \quad
(Fg_{\pm})(k)=\frac{1}{2\textsf{Im}\ k}, \quad \textsf{Im}\ z=2\textsf{Im}\ k\textsf{Re}\ k,
 \end{equation}
and using Lemma \ref{l1} we conclude that the norm of matrix\footnote{the matrix is considered as an operator acting in $\mathbb{C}^2$}
$$
\Phi(k)=\frac{\textsf{Re}\ k}{p_{\mathfrak{T}}(k)}[\mathfrak{T}-\theta_k\mathbf{{det} \ \mathfrak{T}}\sigma_0]
$$
is uniformly bounded on $\mathbb{C}_{+}$. This means that the entries of $\Phi(k)$ must be uniformly bounded
when $k$ runs $\mathbb{C}_{+}$.

Let $A_{\mathfrak{T}}$ has a spectral singularity at $z=\infty$. Then, according to Proposition \ref{neew2},
the polynomial $p_{\mathfrak{T}}(k)$ has no roots. This is possible when $\mathbf{{det} \ \mathfrak{T}}=0$ and
$\gamma_0=0$. In that case $\Phi(k)=\textsf{Re}\ k[\sum_{j=1}^3\gamma_j\sigma_j]$ cannot be uniformly bounded on $\mathbb{C}_+$.
Hence, $A_{\mathfrak{T}}$ is not similar to a self-adjoint operator.

Let $z=0$ be a spectral singularity. Then, $A_{\mathfrak{T}}\not={A_{\frac{1}{2}\sigma_0}}$ and in view of Proposition \ref{neew2},
$p_{\mathfrak{T}}(k)=qk^2$ ($q\not=0$ is some constant). In that case,  at least one of entries of
$\Phi(k)=\displaystyle{\frac{\textsf{Re}\ k}{qk^2}[\mathfrak{T}-\theta_k\mathbf{{det} \ \mathfrak{T}}\sigma_0]}$ tends to infinity when $k\to{0}$. So,
$A_{\mathfrak{T}}$ is not similar to a self-adjoint operator.

Let $z=k_0^2$ be a non-zero spectral singularity. Then $k_0\in\mathbb{R}$ is a root of
$p_{\mathfrak{T}}(k)$ and $\Phi(k)$ tends to infinity when $k\to{k}_0$. Thus
$A_{\mathfrak{T}}$ is not similar to a self-adjoint operator.
Proposition \ref{gg1} is proved. \rule{2mm}{2mm}

{\bf Example IV.} \emph{$\delta'$-potential with a complex coupling.} Let $d\in\mathbb{C}$ and $a=b=c=0$.
Then the expression
$$
 -\frac{d^2}{dx^2}+d<\delta',\cdot>\delta'(x), \qquad d\in\mathbb{C}
$$
determines the operators ${A_d}=\displaystyle{-\frac{d^2}{dx^2}}$ in $L_2(\mathbb{R})$,
which are defined on
$$
 \mathcal{D}(A_d)=\left\{f\in{{W_2^2}(\mathbb{R}\backslash\{0\})} \  \left|\right. \begin{array}{l}
 f'(0+)=f'(0-) \ (\equiv{f'(0)}) \vspace{2mm} \\
 f(0+)-f(0-)=-df'(0)
 \end{array} \right\}
$$
In that case
$$
\mathfrak{T}=\frac{1}{4-2d}\left(\begin{array}{cc}
1-d & 1 \\
1  & 1-d
\end{array}\right), \quad  \mathbf{{det} \ \mathfrak{T}}=-\displaystyle{\frac{d}{2(4-2d)}}, \quad \gamma_0=\displaystyle{\frac{1-d}{4-2d}}.
$$
Substituting these quantities in (\ref{gu8}) we obtain
$$
{\textsf S}(k)=\frac{1}{dk-2i}\left(\begin{array}{cc}
-dk & 2i \\
2i & -dk
\end{array}\right).
$$

The $S$-matrix has a real pole $\displaystyle{k_0=\frac{2i}{d}}$ when $d\in{i}\mathbb{R}\setminus\{0\}$. In that case $\displaystyle{z=k_0^2=\frac{4}{|d|^2}}$
is a spectral singularity of $A_d$.

\setcounter{equation}{0}

\section{Exceptional points}

Let $A$ be a linear operator acting in a Hilbert space $\mathfrak{H}$.
A nonzero vector $f\in\mathcal{D}(A)$ is called \emph{a root vector} of $A$ corresponding to the eigenvalue
$z$ if $(A-zI)^nf=0$ for some $n\in\mathbb{N}$. The set of all roots vectors of $A$ corresponding to a given
eigenvalue $z$, together with zero vector, forms a linear subspace $\mathcal{L}_z$, which is called the \emph{root subspace}. The dimension
of the root subspace $\mathcal{L}_z$ is called the \emph{algebraic multiplicity} of the eigenvalue $z$.  The \emph{geometric multiplicity}
of $z$ is defined as the dimension of the kernel subspace $\ker(A-zI)$ (i.e., as the dimension of the linear subspace of eigenfunctions of $A$ corresponding to $z$).

The algebraic and the geometric multiplicities of $z$  coincide in the case where $A$ is similar to a self-adjoint operator.

The existence of exceptional points deals with
the possible occurrence of nontrivial Jordan blocks in discrete spectra. For
operators  $A_{\mathfrak{T}}$  depending on parameters
$\mathfrak{T}=\{\mathfrak{t}_{ij}\}$  this means that two eigenvalues
$z_1(\mathfrak{T}), \  z_2(\mathfrak{T})$ may coalesce
(degenerate) at certain parameter hypersurfaces of the linear set $\{\mathfrak{t}_{ij}\}$ under
simultaneous coalescence of the corresponding eigenvectors
$f_1(\mathfrak{T}), \ f_2(\mathfrak{T})$ see e.g. \cite{GU}.
We formalize these ideas as follows:

\begin{definition}\label{def1}
Let $A$ be a linear operator acting in a Hilbert space $\mathfrak{H}$.
An eigenvalue $z$ of $A$ is called the exceptional point if the geometric multiplicity
of $z$ does not coincide with its algebraic multiplicity.
\end{definition}

The presence of an exceptional point means that the operator  $A$ is not self-adjoint in $\mathfrak{H}$ and, moreover,
it cannot be self-adjoint for any choice of (equivalent) inner product of $\mathfrak{H}$.

\smallskip

\begin{theorem}\label{new2a}
Let ${\textsf S}(\cdot)$ be the $S$-matrix of
$A_{\mathfrak{T}}$. Then $k_0\in\mathbb{C}_+$  is a pole of order $2$ of ${\textsf S}(\cdot)$ if and only if
$z_0=k^2_0$ is an exceptional point of $A_{\mathfrak{T}}$.
 \end{theorem}
\emph{Proof.}
The resolvent $(A_F-zI)^{-1}$ of a self-adjoint operator $A_F$ (see (\ref{new5}))
is a holomorphic operator-valued function on $\mathbb{C}\setminus\mathbb{R}_+$.

On the other hand, if $A_{\mathfrak{T}}$ is defined by (\ref{sese2c}), then the resolvent $(A_{\mathfrak{T}}-zI)^{-1}$
may be a meromorphic function on $\mathbb{C}\setminus\mathbb{R}_+$ and
an eigenvalue $z_0=k_0^2$ of $A_{\mathfrak{T}}$ will be exceptional if and only if
$(A_{\mathfrak{T}}-zI)^{-1}$ has a pole $z_0$ of order greater than
one\footnote{in our case, the order must be $2$ because
the defect indices of the symmetric operator $A_s$ are $<2,2>$} \cite{WOL}.
Hence, the existence of an exceptional point $z_0=k_0^2$ of $A_{\mathfrak{T}}$ is equivalent
to the existence of pole $z_0$ of order $2$ for the operator-valued function
$$
(A_{\mathfrak{T}}-zI)^{-1}-(A_F-zI)^{-1}.
$$
Taking the proof of Lemma \ref{l1} into account (especially (\ref{new11}) and (\ref{sas89})) we conclude that this condition
is equivalent to the existence of pole $k_0\in\mathbb{C}_+$ of order 2 for the matrix-valued function
${\mathfrak{T}}(\sigma_0-\theta_k{\mathfrak{T}})^{-1}$.

It should be noted that $z_0=-1$ cannot be an exceptional point of $A_{\mathfrak{T}}$ (because $-1\in\rho(A_{\mathfrak{T}})$ for any
operator $A_{\mathfrak{T}}$ defined by (\ref{sese2c})). Hence, the possible pole $k_0\not={i}$ and we
can suppose that $\theta_{k}\not=0$ in some neighbourhood of $\theta_{k_0}=2(1+ik_0)$.
Then
$$
{\mathfrak{T}}(\sigma_0-\theta_k{\mathfrak{T}})^{-1}=-\frac{1}{\theta_k}\sigma_0+\frac{1}{\theta_k}(\sigma_0-\theta_k{\mathfrak{T}})^{-1}.
$$
Comparing the obtained decomposition with (\ref{red1}) we conclude that
$k_0$ is pole of order 2 of ${\mathfrak{T}}(\sigma_0-\theta_k{\mathfrak{T}})^{-1}$
if and only if $k_0$ is pole of order 2 of the $S$-matrix ${\textsf S}(\cdot)$.
Theorem \ref{new2a} is proved.
\rule{2mm}{2mm}

\begin{corollary}\label{new14}
The point $z_0=k^2_0$ is an exceptional point of $A_{\mathfrak{T}}$
if and only if the matrix $\sigma_0-\theta_{k_0}\mathfrak{T}$ is nonzero and nilpotent.
\end{corollary}
\emph{Proof.}  Let $z_0=k^2_0$ be an exceptional point of $A_{\mathfrak{T}}$. Then
$k_0\in\mathbb{C}_+$ is a pole of order 2 for ${\textsf S}(\cdot)$.
Taking (\ref{gu8}) into account, we conclude that $\mathbf{{det} \ \mathfrak{T}}\not=0$ and $k_0\not=i$.
So, ${\textsf S}(\cdot)$ is defined by (\ref{gu8c}) and
$\theta_{k_{0}}=2(1+ik_0)\not=0$.

The $S$-matrix ${\textsf S}(\cdot)$ has a pole $k_0$ of order 2 if and only if at least one of
functions $s_j(\cdot)$ in the decomposition (\ref{new3}) has pole $k_0$ of order 2.
In that case, the simple analysis of (\ref{au1}) shows that
$\theta_+=\theta_-=\theta_{k_0}$. Then, in view of (\ref{j5}),
\begin{equation}\label{j1}
\sum_{j=1}^3{\gamma_j^2}=0, \qquad \theta_+=\theta_-=\theta_{k_0}=\frac{1}{\gamma_0}.
\end{equation}
We note that not all coefficients $\gamma_j$ are equal to zero in the first relation of (\ref{j1}).
 Indeed, suppose that $\gamma_1=\gamma_2=\gamma_3=0$.
Then $\mathfrak{T}=\gamma_0\sigma_0$ and $\theta_-\theta_+=1/\gamma_0^2$. Substituting these quantities into (\ref{gu8c}), we obtain
$$
{\textsf S}(k)=\left[1-\frac{4ik}{\theta_k-{1}/{\gamma_0}}\right]\sigma_0=-\frac{k+k_0}{k-k_0}\sigma_0.
$$
Therefore, $k_0$ cannot be a pole of order $2$. The obtained contradiction means that at least one of coefficients $\gamma_1, \gamma_2, \gamma_3$ differs from zero. In that case, the matrix
$$
\sigma_0-\theta_{k_0}\mathfrak{T}=(1-\theta_{k_0}\gamma_0)\sigma_0-\theta_{k_0}\sum_{j=1}^{3}\gamma_j\sigma_j=-\theta_{k_0}\sum_{j=1}^{3}\gamma_j\sigma_j
$$
is nonzero.

On the other hand, taking (\ref{j1}) and properties of Pauli matrices into account,
$$
(\sigma_0-\theta_{k_0}\mathfrak{T})^2=\theta_{k_0}^2\left(\sum_{j=1}^{3}\gamma_j\sigma_j\right)^2=\theta_{k_0}^2\left(\sum_{j=1}^3{\gamma_j^2}\right)\sigma_0=0.
$$

Conversely, let $\sigma_0-\theta_{k_0}\mathfrak{T}$ be a nonzero and nilpotent matrix.
In that case
$$
\sigma_0-\theta_{k_0}\mathfrak{T}=(1-\theta_{k_0}\gamma_0)\sigma_0-\theta_{k_0}\sum_{j=1}^{3}\gamma_j\sigma_j\not={0}
$$
and
$$
0=(\sigma_0-\theta_{k_0}\mathfrak{T})^2=\left[(1-\theta_{k_0}\gamma_0)\sigma_0-\theta_{k_0}\sum_{j=1}^{3}\gamma_j\sigma_j\right]^2=
$$
$$
[(1-\theta_{k_0}\gamma_0)^2+\theta_{k_0}^2\sum_{j=1}^3\gamma_j^2]\sigma_0+2(1-\theta_{k_0}\gamma_0)\theta_{k_0}\sum_{j=1}^3\gamma_j\sigma_j.
$$
These relations are possible only in the case: $1-\theta_{k_0}\gamma_0=0$ and $\sum_{j=1}^3\gamma_j^2=0$, where at least one
$\gamma_j$ differs from zero.
Then $\theta_{k_0}=\theta_+=\theta_-=\frac{1}{\gamma_0}$ and $k_0=i-\frac{i}{2\gamma_0}$ is a pole of order 2 of
${\textsf S}(\cdot)$. Corollary \ref{new14} is proved.
\rule{2mm}{2mm}

In the particular case where  $A_{\mathfrak{T}}=A_{\mathbf{T}}$, the (possible) appearance of
exceptional point is determined by parameters $a, d$.

\begin{cor}\label{bbb9}
Let $z_0=k^2_0$ be an exceptional point of $A_{\mathbf{T}}$.
Then
\begin{equation}\label{j10}
k_0=\displaystyle{-i\frac{4-\mathbf{{det} \ T}+4a}{4-\mathbf{{det} \ T}-4d}}=i\frac{4-\mathbf{{det} \ T}}{4d}
\end{equation}
and $z=k_0^2=\displaystyle{\frac{a}{d}}$.
\end{cor}
\emph{Proof.}
If $z_0=k^2_0$ is an exceptional point of $A_{\mathbf{T}}$, then $k_0\in\mathbb{C}_+$  is a pole of order $2$ of ${\textsf S}(\cdot)$.
Then $k_0=i-\frac{i}{2\gamma_0}$ and
the first relation in (\ref{j10}) follows from (\ref{gu9}).

Using (\ref{gu10}), (\ref{j1}) and taking into account that
$$
(4-\mathbf{{det} \ T})^2+16ad=(4-ad+bc)^2+16ad=(4+ad-bc)^2+16bc=(4+\mathbf{{det} \ T})^2+16bc
$$
we conclude that $(4-\mathbf{{det} \ T})^2+16ad=0$.
Therefore,
$$
k_0=-i\frac{4-\mathbf{{det} \ T}+4a}{4-\mathbf{{det} \ T}-4d}=-i\frac{4-\mathbf{{det} \ T}+4a}{4-\mathbf{{det} \ T}-4d}\cdot\frac{4-\mathbf{{det} \ T}+4d}{4-\mathbf{{det} \ T}+4d}=i\frac{4-\mathbf{{det} \ T}}{4d}.
$$
To complete the proof it suffices to calculate
$$
z=k_0^2=-\frac{(4-\mathbf{{det} \ T})^2}{16d^2}=\frac{16ad}{16d^2}=\frac{a}{d}.
$$
\rule{2mm}{2mm}

The S-matrices in Examples I-IV do not have poles of order $2$. Hence, the corresponding operators
$A_{\mathfrak{T}}$ do not have exceptional points.

{\bf Example V.}  Let $a=-e^{i\phi}$, $b=-1$, $c=1$, and $d=e^{-i\phi}$. Then (\ref{lesia11}) takes the form
$$
 -\frac{d^2}{dx^2}-e^{i\phi}<\delta,\cdot>\delta(x)-<\delta',\cdot>\delta(x)+
 <\delta,\cdot>\delta'(x)+e^{-i\phi}<\delta',\cdot>\delta'(x)
$$
and (\ref{lesia800}) determines the operators ${A_\phi}=\displaystyle{-\frac{d^2}{dx^2}}$ with domains of definition
$$
 \mathcal{D}(A_\phi)=\left\{f\in{{W_2^2}(\mathbb{R}\backslash\{0\})} \  \left|\right.  \begin{array}{l}
 f(0+)+e^{-i\phi}f'(0+)=2f(0-)  \vspace{2mm} \\
 f(0-)=e^{-i\phi}f'(0-)  \end{array} \right\}.
 $$

If $\phi\in(0, 2\pi)$, then ${\mathcal D}(A_{\phi})$ can be presented
in the form (\ref{sese2c}), where
$$
\mathfrak{T}=\frac{1}{8\sin^2\phi/2}\left(\begin{array}{cc}
1-e^{-i\phi} & 2 \\
0  & 1-e^{-i\phi}
\end{array}\right).
$$
In that case
$$
\theta_-=\theta_+=2(1-e^{i\phi}),  \qquad  \mathbf{{det} \ \mathfrak{T}}=\frac{1}{4(1-e^{i\phi})}\not=0.
$$

Substituting these quantities in (\ref{gu8c}) we obtain
$$
{\textsf S}(k)=-\frac{k+ie^{i\phi}}{k-ie^{i\phi}}\sigma_0+ \frac{2ik}{(k-ie^{i\phi})^2}\left(\begin{array}{cc}
0 & 2e^{i\phi} \\
0  & 0
\end{array}\right).
$$

The $S$-matrix has pole $k_0=ie^{i\phi}$ of order $2$ in the physical sheet $\mathbb{C}_+$ when
$\phi\in[0, \frac{\pi}{2})\cup(\frac{3\pi}{2},2\pi]$. In that case $z_0=-e^{2i\phi}$ is the exceptional point of $A_{\phi}$.

\smallskip

If $\phi$ coincides with $\frac{\pi}{2}$ or with  $\frac{3\pi}{2}$, then the $S$-matrix has real poles $k_0=-1$ or $k_0=1$, respectively.
The operator $A_{\phi}$ has spectral singularity $z_0=1$.

\smallskip

If $\phi\in(\frac{\pi}{2}, \frac{3\pi}{2})$, then the pole $k_0$ of the $S$-matrix belongs to the nonphysical sheet
$\mathbb{C}_-$. The corresponding operator $A_\phi$ is similar to self-adjoint.

\section{Conclusions}
This paper shows that poles of $S$-matrix ${\textsf S}(\cdot)$ completely characterize the properties of
Schr\"{o}dinger operators $A_{\mathfrak{T}}$ with non-symmetric zero-range potentials (\ref{gu14}).
Precisely, poles of ${\textsf S}(\cdot)$ on the physical sheet $\mathbb{C}_+$  describe the discrete spectrum $\sigma_p$ of
$A_{\mathfrak{T}}$. The appearance of exceptional points on $\sigma_p$ is distinguished by poles of order $2$
on the physical sheet. The existence of spectral singularities on the continuous spectrum $\sigma_c$ of
$A_{\mathfrak{T}}$ is determined by poles of ${\textsf S}(\cdot)$ on the extended real line $\overline{\mathbb{R}}=\mathbb{R}\cup\{\infty\}$.
The property of similarity of $A_{\mathfrak{T}}$ to a self-adjoint operator means that the $S$-matrix
${\textsf S}(\cdot)$ has poles in the nonphysical sheet $\mathbb{C}_-$ or ${\textsf S}(\cdot)$
has simple non-zero imaginary poles.

Not every operator $A_{\mathfrak{T}}$ defined by (\ref{sese2c}) and studied in the paper can be interpreted as pseudo-hermitian or $\mathcal{PT}$-symmetric. Sometimes \cite{AG}, such more general class of operators is called \emph{quasi-self-adjoint}.
Our studies show that techniques based on the decomposition of $S$-matrix with respect to the Pauli matrices
have proved very useful for investigation of quasi-self-adjoint operators.
In this way, we find an explicit expression of metric operators for the case where
$S$-matrix has simple non-zero imaginary poles.

The results of the paper were established with the use of expression (\ref{red1}) for $S$-matrices which comes from the Lax-Phillips scattering
theory and it is closed to the concept of characteristic function of quasi-self-adjoint operators \cite{AG}. Using the equivalent
representation (\ref{rest2}) of the $S$-matrix we can reformulate the obtained results in terms of reflection and transmission coefficients.

The methods developed in the paper can be applied for studies of $S$-matrices of Schr\"{o}dinger operators with non-symmetric potentials
having a compact support.

\section{Acknowledgments}
A. Grod is supported in part by a grant
no. 03-01-12 of National Academy of Sciences of Ukraine.

\end{document}